\documentclass[pre,twocolumn,showpacs]{revtex4-1}

\usepackage{graphicx}
\usepackage[latin1]{inputenc}
\usepackage{amsfonts}
\usepackage{amssymb}
\usepackage{amsmath,amssymb}
\usepackage{bm}
\usepackage{cancel}
\usepackage{cleveref}
\usepackage{subfigure} 
\usepackage{bbold}
\usepackage{caption}
\usepackage{color}

\begin{document}
\graphicspath{{images/}}

\title{Charging capacitors using diodes at different temperatures. I Theory}
\author{L. L. Bonilla$^*$}
\affiliation{Universidad Carlos III de Madrid, ROR: https://ror.org/03ths8210, Departamento de Matem\'aticas, Avenida de la Universidad, 30 (edificio Sabatini), 28911 Legan\'es (Madrid), Spain} 
\affiliation{Universidad Carlos III de Madrid,  ROR: https://ror.org/03ths8210, G. Mill\'an Institute, Fluid Dynamics, Nanoscience and Industrial Mathematics, Avenida de la Universidad 30 (edificio Sabatini); 28911 Legan\'{e}s, Spain
$^*$Corresponding author. E-mail: bonilla@ing.uc3m.es}
\author{A. Torrente}
\affiliation{Universidad Carlos III de Madrid, ROR: https://ror.org/03ths8210, Departamento de Matem\'aticas, Avenida de la Universidad, 30 (edificio Sabatini), 28911 Legan\'es (Madrid), Spain} 
\affiliation{Universidad Carlos III de Madrid,  ROR: https://ror.org/03ths8210, G. Mill\'an Institute, Fluid Dynamics, Nanoscience and Industrial Mathematics, Avenida de la Universidad 30 (edificio Sabatini); 28911 Legan\'{e}s, Spain}

\author{J. M. Mangum}
\affiliation{Department of Physics, University of Arkansas, Fayetteville, Arkansas 72701, USA}
\author{P. M. Thibado}
\affiliation{Department of Physics, University of Arkansas, Fayetteville, Arkansas 72701, USA}
\date{\today}

\begin{abstract}
Nonlinear elements in a rectifying circuit can be used to harvest energy from thermal fluctuations either steadily or transitorily. We study an energy harvesting system comprising a small variable capacitor (e.g., free standing graphene) wired to two diodes and two storage capacitors that may be kept at different temperatures (or at a single one) and use two current loops. The system reaches very rapidly a quasi stationary state with constant overall charge while the difference of the charges at the storage capacitors evolves much more slowly to its stationary value. In this paper, we extract an exponentially small factor out of the solution of the Fokker-Planck equation and use a Chapman-Enskog procedure to describe the long evolution of the marginal probability density for the charge difference, from the quasi stationary state to the final stationary state (thermal equilibrium for equal temperatures). The second paper of this series shows that the results of the perturbation procedure compare well with direct numerical simulations. For a specific form of the diodes' nonlinear mobilities, we can approximate the quasi stationary state by Gaussian functions and further study the evolution of the marginal probability density. The latter adopts the shape of a slowly expanding pulse (comprising left and right moving wave fronts whose fore edges become sharper as time elapses) in the space of charge differences that leaves the final stationary state behind it. 
\end{abstract}

\maketitle

\section{Introduction}\label{sec:1}
While challenging, extracting energy from thermal baths using a small engine could be used to power small devices (even nanodevices \cite{fil07,vio24,mur25}). There are electrical devices that consume little (nanowatts in active mode and picowatts in standby mode \cite{ban16,han09,lee13,bas22,gup16}). This amount of power can easily be found from electromagnetic sources and even from mechanical vibrations in noisy environments \cite{cos24}. When fluctuations are dominant, optimization protocols need to be applied to engines \cite{aur11}. Energy harvesting in a quiet, dark setting is the most challenging because only thermal energy is present. In such an environment, the Brownian motion of electrons produces a stochastic alternating current at a single temperature and a natural question is: {\em Can one extract energy out of thermal fluctuations?} The answer is no, for the second law of thermodynamics precludes this in presence of a single thermal bath: The system will eventually go to thermal equilibrium. Using a diode to rectify an alternating current produced by the Brownian motion of electrons to charge a capacitor was shown to violate detailed balance by Brillouin \cite{bri50}. The impossibility to harvest thermal energy at a single temperature was discussed by Feynman in his Ratchet and Pawl lecture about a mechanical rectifier \cite{feynman}. There he also argued that the efficiency of his system as a thermal engine with ratchet and pawl at different temperatures will be at most the Carnot one, a point corrected by later studies \cite{mag98}. 
 
Detailed analyses of these systems are carried out using Fokker-Planck equations (FPEs) to characterize the stationary states to which they evolve \cite{mag98,vka60,lan62,sok98,sok99}. However, we know that going to thermal equilibrium may take a very long time due to the presence of nonlinear elements, during which transient states may produce useful outcomes \cite{thi23}. It is also interesting to consider the time these systems take to reach the stationary nonequilibrium state maintained by thermal gradients. In one such system, the key elements are a fluctuating freestanding graphene sheet (that acts as a small variable capacitor) coupled to a rectifying circuit with two nonlinear diodes and two storage capacitors via the tip of a scanning tunneling microscope (STM) \cite{thi23}. 
 
\begin{figure}[h]
\begin{center}
\includegraphics[width=7cm]{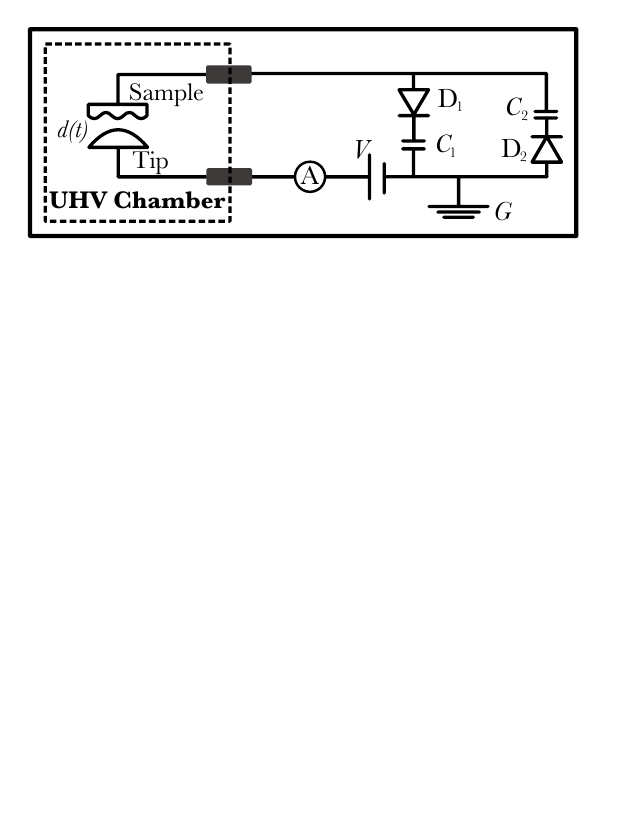}
\end{center}
\caption{Circuit diagram showing the STM tip and sample equivalent to a small capacitor $C_0$, and the opposing diodes D1 and D2, with respective conductances $\mu_1$, $\mu_2\propto R^{-1}$, and storage capacitors $C_1$ and $C_2$. The current-voltage curve of each diode is similar to that of an ideal diode in series with a resistor $R$, and therefore resistances are included in the diodes. \label{fig1}}
\end{figure}

 As shown in Fig.~\ref{fig1}, the graphene fluctuations cause the distance $d(t)$ to the STM tip to change with time. Thus, the freestanding graphene acts as a capacitor of variable capacitance $\varepsilon A/d(t)$ (permittivity $\varepsilon$, effective area $A$) sending the generated displacement current to one diode or the other depending on its sign and charging the storage capacitor attached to it. If the whole graphene and circuit system is kept at a single temperature, there is temporary charging of the capacitors and the time it takes to discharge them and reach thermal equilibrium may be very large depending on the ratio between the capacitances of graphene and the storage capacitors, and on the nonlinearity of the diodes \cite{thi23}. Charging the storage capacitors may take a very short time while the discharging stage is very long \cite{thi23}. Then the charged capacitors can be disconnected and their stored energy used to power electrical devices. If we keep the diodes at different temperatures, the overall system reaches a stationary state from which it is possible to extract work on a steady basis.
  
The analysis of such systems poses problems typical of singular perturbations \cite{bender,neu}. The small capacitance ratio causes terms in the energy of the circuit to be of vastly different magnitude, which then appear exponentially in the equilibrium canonical probability density if the overall system is at a single temperature. The system can be extended to a number of similar units connected in parallel to the variable capacitor. Describing the transient stage requires exponential asymptotics and a method to deduce the evolution of the system at the slow time scale towards equilibrium. The main idea is to factor out the equilibrium state from the probability density that solves the FPE and then using a Chapman-Enskog expansion \cite{bon10,bon19} to approximate the resulting equation. 

In the first paper of this series, we present the analysis of the energy harvesting system comprising a small variable capacitor wired to two diodes and two storage capacitors that are kept at different temperatures and use two current loops. The system evolves rapidly to a quasistationary state in which the average total charge is zero times a slowly varying marginal probability of the charge difference between the capacitors. The latter evolves very slowly to the final stationary state adopting the shape of an expanding pulse (comprising left and right moving wave fronts) in the space of charge differences that leaves the final stationary state behind it. The evolution time increases exponentially with the position of the wave fronts. While this general picture follows from Chapman-Enskog asymptotics for the ratio of the probability density to that of the equilibrium state, it is possible to calculate the coefficient functions of the reduced equation for a specific form of the diode mobility close to a step function. Using this additional approximation, we show that the profiles of the wave fronts are displaced Gaussian functions whose variances decrease as the fronts advance. It turns out that the front thicknesses decrease faster in the approach to equilibrium when the diode temperatures are the same, compared to the approach to the nonequilibrium stationary state produced by a temperature difference.

The second paper of this series presents direct numerical simulations of the FPE for the probability density of the energy harvesting system and compares them to the numerical solutions of the Chapman-Enskog method. The perturbation results approximate well those of direct numerical simulations within a wide range of parameters.

The rest of this paper is as follows. Section \ref{sec:2} reviews the Fokker-Planck description of the harvesting system, its nondimensional form, the equilibrium probability density for the case of a single temperature and a simplified version of the stationary state when the two diodes are at different temperatures. Section \ref{sec:3} presents the basis of our approach. First, the equations for the ratio of the time dependent density to the equilibrium state for the average temperature, second the Chapman-Enskog method and the reduced equation for the marginal probability density. The details of the derivations are given in Appendices \ref{ap:a} and \ref{ap:b}. Section \ref{sec:4} discusses the stationary state and different approximations that hold for particular versions of the diode mobility. For them, the coefficients appearing in the reduced equation can be approximated as explained in Appendix \ref{ap:c} and the stationary averages and variances are calculated in Appendix \ref{ap:d}. The evolution of the marginal probability density to the final stationary state is considered in Section \ref{sec:5}. We describe the profile of the marginal probability density in the space of the difference of charge at the storage capacitors: its shape as an expanding pulse that leaves inside the final stationary state. Details of the calculations are given in Appendix \ref{ap:d}. Section \ref{sec:6} discusses our results. Appendix \ref{ap:e} discusses results for piecewise linear diode mobilities.

\section{Fokker-Planck equation and equilibrium}\label{sec:2}
Consider the system of Fig.~\ref{fig1} without battery, $V=0$, and with the diodes kept at different temperatures $T_1$ and $T_2$. The freestanding graphene fluctuates at a much faster rate than the time scale of the current at the circuit. Then it is at local equilibrium with the instantaneous value of the charges in the circuit (see Appendix A of \cite{thi23}) and it acts as a capacitor with average capacitance $C_0=\varepsilon A/d$, where $\varepsilon$ is the permittivity, $A$ is the effective area, and $d$ is the fixed distance between the membrane support and the STM tip. The probability density of having charges $q_1$ and $q_2$ at the storage capacitors satisfies the following FPE \cite{thi23}:
\begin{subequations}\label{eq1}
\begin{eqnarray}
&&\frac{\partial\rho}{\partial t}-\sum_{i=1}^2\!\frac{\partial}{\partial q_i}\!\left[\mu_i(u_i)\left(\rho \frac{\partial\mathcal{H}}{\partial q_i}+k_BT_i \frac{\partial\rho}{\partial q_i}\right)\right]\!=0, \label{eq1a}\\
&& \int\rho(q_1,q_2,t)\, dq_1dq_2=1,\label{eq1b}\\
&&\mathcal{H}=\frac{(q_1+q_2)^2}{2C_0}+\sum_{j=1}^2\frac{q_j^2}{2C_1},\quad u_i=-\frac{\partial\mathcal{H}}{\partial q_i},      \label{eq1c}\\
&&\mu_i(u)=\mu((-1)^{i+1}u), \quad\mu(u)=\frac{1}{R}\frac{1}{1+e^{-u/u_0}}. \label{eq1d}
\end{eqnarray}\end{subequations}
Here $\mathcal{H}(q_1,q_2)$ is the electrostatic energy, $u_i$ is the voltage across the capacitor associated to diode $i$, $i=1,2$,  $C_1$ is the common capacitance of the two identical storage capacitors, and $\mu(u)$ is the conductance of the two identical diodes. When the diodes are oriented as in Fig.~\ref{fig1}, their conductances are $\mu_1=\mu(u)$ and $\mu_2(u)=\mu(-u)$. In Eq.~\eqref{eq1d}, $u$ is the voltage across the diode, $R$ is the resistance at high forward bias, and $u_0$ sets the quality of the diode. For example, as $u_0$ approaches zero, the diode conductance matches that of an ideal switch, and as $u_0$ becomes larger the diode conductance becomes a linear resistor. The sigmoid function $\mu(u)$ mimics an ideal diode in series with a resistor, which more closely represents a real diode \cite{sze81}.

Starting from an initial state of zero charge in the circuit, the average charges at the storage capacitors increase rapidly and then relax very slowly to their final stationary value. From  the FPE (\ref{eq1a}), the current through capacitor $i$ is:
\begin{eqnarray*}
\frac{d}{dt}\langle q_i \rangle = \langle u_i \mu_i(u_i)\rangle - k_B T _i\!\left(\frac{1}{C_0} + \frac{1}{C_1}\right) \langle\mu'_i(u_i) \rangle. 
\end{eqnarray*}
This is the sum of the usual Ohm's law current and the nonlinear thermal current proportional to the temperature $T_i$. If $\rho = \delta(q_1) \delta(q_2)$ initially, $\left. \frac{d}{dt}\langle q_i \rangle\right|_{t=0} = - k_B T_i (\frac{1}{C_0} + \frac{1}{C_1}) \mu'_i(0)$ and the capacitors charge due to their nonzero conductance slope $\mu'_i(0)= (-1)^{i+1}/(4Ru_0)$, which can be large for small $u_0$.  The initial diode current puts positive charges on storage capacitor associated to diode $D_1$ and negative charges on the capacitor associated to diode $D_2$ of Fig.~\ref{fig1}. A detailed discussion of initial charging is given in Appendix \ref{ap:a}.

After the initial stage, the system evolves on a slower time scale to its final stationary state. To describe this stage, it is convenient to rewrite the FPE \eqref{eq1} in nondimensional units that involve the sum and difference of the charges $q_1$ and $q_2$ as variables:
\begin{subequations}\label{eq2}
\begin{eqnarray}
&&\xi=\epsilon\frac{q_1\!-\!q_2}{C_0V_0}=\frac{q_1\!-\!q_2}{2C_1V_0}, \,\eta=\frac{q_1\!+\!q_2}{C_0V_0}(1\!+\!\epsilon),\, \tau\!=\!\frac{t}{2RC_1},\quad\label{eq2a}\\
&&\epsilon=\frac{C_0}{2C_1}, \quad  V_0=\sqrt{\frac{k_B(T_1+T_2)}{2C_0}},\quad w=\frac{u_0}{V_0}.\label{eq2b}
\end{eqnarray}\end{subequations}
Typically $\epsilon\ll 1$ and $w\ll 1$. Note that decreasing the resistance $R$ by a numerical factor $N>0$ decreases the $RC_1$ time scale by the same factor according to Eq.~\eqref{eq2a}. This will be demonstrated in the second paper of this series by placing $N$ identical diode-capacitor pairs in parallel and observing the shortening of times in the charging dynamics. 

Substituting Eqs.~\eqref{eq2} into Eq.~\eqref{eq1}, we obtain the following nondimensional FPE: 
\begin{widetext}
\begin{subequations}\label{eq3}
\begin{eqnarray}
&&\epsilon\frac{\partial \rho}{\partial\tau}+\frac{\partial J_\eta}{\partial \eta} + \frac{\partial J_\xi}{\partial\xi}=0, \quad \int\rho(\eta,\xi,\tau)\, d\eta\, d\xi=1,\label{eq3a}\\
&&\!\! J_\eta=(1+\epsilon)\!\left[(\mu_1 \!+\mu_2)\eta+(\mu_1\!-\mu_2)\xi+(1+\epsilon)(\theta_1\mu_1\!+\theta_2\mu_2)\frac{\partial}{\partial\eta}+\epsilon(\theta_1\mu_1-\theta_2 \mu_2)\frac{\partial}{\partial\xi}\right]\!\rho,\quad\,\, \label{eq3b}\\
&& J_\xi=\epsilon\left[(\mu_1-\mu_2)\eta+(\mu_1+\mu_2)\xi+(1+\epsilon)(\theta_1\mu_1-\theta_2\mu_2)\frac{\partial}{\partial\eta}+\epsilon(\theta_1\mu_1+\theta_2 \mu_2) \frac{\partial}{\partial\xi}\right]\!\rho, \label{eq3c}\\
&&\theta_1=\frac{2T_1}{T_1+T_2}, \quad\theta_2=\frac{2T_2}{T_1+T_2}, \quad \theta_1+\theta_2=2,\label{eq3d}\\
&&\quad\mu_i= \mu(-\eta+(-1)^i\xi),\,\, i=1,2,\quad \mu(u)=\frac{1}{1+e^{-u/w}}. \label{eq3e}
\end{eqnarray}\end{subequations}
\end{widetext}
Here $J_\eta$ and $J_\xi$ are the components of the probability current density. The initial condition for Eq.~\eqref{eq3a} comes from the initial charging state and is proportional to $\delta(\xi)$; see Eq.~\eqref{eqa7}. At the stationary state, the divergence of the probability current vector is zero. 

For a single temperature, $\theta_1=\theta_2=1$, the stationary state is the equilibrium state with probability density  
\begin{subequations}\label{eq4}
\begin{eqnarray}
\rho_\text{eq}(\eta,\xi)= \frac{1}{2\pi\sqrt{\epsilon(1+\epsilon)}}\exp\!\left(-\frac{\eta^2}{2(1+\epsilon)}-\frac{\xi^2}{2\epsilon}\right)\!, \label{eq4a}
\end{eqnarray}
corresponding to the nondimensional energy
\begin{eqnarray}
\mathcal{H}=\frac{\eta^2}{2(1+\epsilon)}+\frac{\xi^2}{2\epsilon}. \label{eq4b}
\end{eqnarray}
\end{subequations}
Notice that Eqs.~\eqref{eq4} corresponds to thermal equilibrium at the average temperature $(T_1+T_2)/2$ that enters the potential $V_0$ of Eq.~\eqref{eq2b}. The two terms in the energy \eqref{eq4b} have orders 1 and $1/\epsilon\gg 1$ provided the dimensionless charges $\eta$ and $\xi$ are of order 1. This is the origin of the exponential asymptotics which is the basis of our analysis in the present paper. Note that the equilibrium density \eqref{eq4a} becomes proportional to $\delta(\xi)$ as $\epsilon\to 0+$. Assuming this to be the case for the stationary state when the diode temperatures are different (we will get better results later), integrating Eq.~\eqref{eq3b} over $\xi$ and equating the result to 0, we obtain the approximate stationary state: 
\begin{subequations}\label{eq5}
\begin{eqnarray}
&&\rho_s(\eta,0)=\frac{1}{Z}\, \exp\!\left[-\frac{1}{1+\epsilon}\int \frac{(\mu_2+\mu_1)\eta}{\theta_1\mu_1+\theta_2\mu_2}\, d\eta \right]\!,\label{eq5a}\\
&&Z=\int_{-\infty}^\infty \exp\!\left[-\frac{1}{1+\epsilon}\int \frac{(\mu_2+\mu_1)\eta}{\theta_1\mu_1+\theta_2\mu_2}\, d\eta \right]d\eta.    \label{eq5b}
\end{eqnarray}
Going back to dimensional variables, this is Eq. (3) in \cite{sok99} for the case of a single capacitor with charge $q$, voltage drop $u=q/C$, and capacitance $C=(1+\epsilon)/C_0$ that is coupled to two diodes. With units, the average total charge and variance are
\begin{eqnarray}
&&\!\langle q\rangle\!\!= \!CV_0\!\langle\eta\rangle\!\!=\!\frac{CV_0}{Z}\!\!\int_{-\infty}^\infty\!\eta \exp\!\!\left[-\frac{1}{1\!+\!\epsilon}\!\!\int\!\! \frac{\eta(\mu_1\!+\!\mu_2)d\eta}{\theta_1\mu_1\!+\!\theta_2\mu_2}\! \right]\!\!d\eta,\quad \label{eq5c}\\
&&\langle(\Delta q)^2\rangle=(CV_0)^2(\langle\eta^2\rangle-\langle\eta\rangle^2).\label{eq5d}
\end{eqnarray}
\end{subequations}

\section{Chapman-Enskog derivation of the reduced FPE}\label{sec:3}
The initial charging state occurs on the fast time scale $\tau/\epsilon$ and it is described in Appendix \ref{ap:a}. After this fast phase, the charges slowly relax to their stationary values. We expect the description of this last stage to be rather subtle for the final state should become the singular expression \eqref{eq4a} if $\theta_j=1$. To make sure that the final stationary state is compatible with thermal equilibrium, we extract the factor $\rho_\text{eq}$ of Eq.~\eqref{eq4a} (equilibrium at the average temperature $(T_1+T_2)/2$) from the probability density,
\begin{subequations}\label{eq6}
\begin{eqnarray}
\rho(\eta,\xi,\tau)= \rho_\text{eq}(\eta,\xi)\, \tilde{\rho}(\eta,\xi,\tau), \label{eq6a}
\end{eqnarray}
and write Eq.~\eqref{eq3a} in terms of $\tilde{\rho}$:
\begin{eqnarray}
\epsilon\rho_\text{eq}\frac{\partial\tilde{\rho}}{\partial\tau}+\frac{\partial \tilde{J}_\eta}{\partial \eta} + \frac{\partial \tilde{J}_\xi}{\partial\xi}=0.\label{eq6b}
\end{eqnarray}
\end{subequations}
From Eqs.~\eqref{eq3} and \eqref{eq6a}, we extract the probability currents that appear in Eq.~\eqref{eq6b}:
\begin{widetext}
\begin{subequations}\label{eq7}
\begin{eqnarray}
\tilde{J}_\eta\!&=&(1+\epsilon)\rho_\text{eq}\left[\frac{\theta_1-\theta_2}{2}[(\mu_2-\mu_1)\eta-(\mu_1+\mu_2)\xi]+(1+\epsilon)(\theta_1\mu_1+\theta_2\mu_2)\frac{\partial}{\partial\eta}\right.\nonumber\\
\!&+&\!\left.\epsilon(\theta_1\mu_1-\theta_2 \mu_2)\frac{\partial}{\partial\xi}\right]\!\tilde{\rho}, \label{eq7a}\\
\tilde{J}_\xi\!&=&\!\epsilon\rho_\text{eq}\left[\frac{\theta_1-\theta_2}{2}[(\mu_2-\mu_1)\xi-(\mu_1+\mu_2)\eta]+(1+\epsilon)(\theta_1\mu_1-\theta_2\mu_2)\frac{\partial}{\partial\eta}\right.\nonumber\\
\!&+&\!\left.\epsilon(\theta_1\mu_1+\theta_2 \mu_2) \frac{\partial}{\partial\xi}\right]\!\tilde{\rho}. \label{eq7b}
\end{eqnarray}\end{subequations}

Eq.~\eqref{eq6b} together with Eqs.~\eqref{eq7} can be written as
\begin{subequations}\label{eq8}
\begin{eqnarray}
&&\epsilon\rho_\text{eq}\frac{\partial\tilde{\rho}}{\partial \tau}=(\mathcal{L}+ \epsilon\mathcal{N}_1+\epsilon^2\mathcal{N}_2)\tilde{\rho},  \label{eq8a}\\
&&\mathcal{L}\!=(1+\epsilon)\frac{\partial}{\partial\eta}\!\left\{\rho_\text{eq}\left[\frac{\theta_1-\theta_2}{2}[(\mu_2\!-\!\mu_1)\eta-(\mu_1+\mu_2)\xi]+(1+\epsilon)(\theta_1\mu_1+\theta_2\mu_2) \frac{\partial}{\partial \eta} \right]\!\right\}\!,\quad \quad\label{eq8b}\\
&&\mathcal{N}_1\!= (1+\epsilon)\frac{\partial}{\partial\eta}\!\left[\rho_\text{eq} (\theta_1\mu_1\!-\!\theta_2\mu_2) \frac{\partial}{\partial\xi}\!\right]\!+\nonumber\\
&&\quad\,+ \frac{\partial}{\partial\xi}\!\left\{\rho_\text{eq} \left[\frac{\theta_1-\theta_2}{2}[(\mu_2\!-\!\mu_1)\xi-(\mu_1+\mu_2)\eta]+(1+\epsilon)(\theta_1\mu_1-\theta_2\mu_2) \frac{\partial}{\partial \eta} \right]\!\right\}\! , \quad  \label{eq8c}\\
&&\mathcal{N}_2=\frac{\partial}{\partial\xi}\!\left[\rho_\text{eq}(\theta_1\mu_1+\theta_2\mu_2) \frac{\partial}{\partial\xi}\right]\!.     \label{eq8d}
\end{eqnarray}\end{subequations}
\end{widetext}

\noindent We consider $\mathcal{L}$, $\mathcal{N}_1$ and $\mathcal{N}_2$ to be of order 1 as $\epsilon\to 0$, notwithstanding the $\epsilon$-dependance of the equilibrium density and having left the factors $(1+\epsilon)$ unchanged.  For the remainder of this section, we shall not use the specific form \eqref{eq3e} of the mobilities. 

The leading order equation $\mathcal{L}\tilde{\rho}=0$ has the following solution  with zero current at $\eta=\pm\infty$:
\begin{subequations}\label{eq9}
\begin{eqnarray}
&&\tilde{\rho}^{(0)}(\eta,\xi)= \exp\!\left[-\frac{(\theta_1-\theta_2)\mathcal{J}(\eta,\xi)}{2(1+\epsilon)}\right] Q^{(0)}(\xi,\tau),\label{eq9a}\\
&&\mathcal{J}(\eta,\xi)=\int_0^\eta \frac{(\mu_2-\mu_1)\eta-(\mu_1+\mu_2)\xi}{\theta_1\mu_1+\theta_2\mu_2}\, d\eta.\quad\label{eq9b}
\end{eqnarray}
The corresponding approximate probability density is
\begin{eqnarray}
&&\rho^{(0)}(\eta,\xi;P)=\hat{E}(\eta,\xi)P(\xi,\tau), \label{eq9c}\\
&&P(\xi,\tau)= e^{-\frac{\xi^2}{2\epsilon}}Q(\xi,\tau),\quad\int_{-\infty}^\infty P\, d\xi=1,\label{eq9d}\\
&&\hat{E}(\eta,\xi)=\frac{E(\eta,\xi)}{\overline{E(\eta,\xi)}},\quad E(\eta,\xi)= e^{-\frac{\eta^2+(\theta_1-\theta_2)\mathcal{J}\!(\eta,\xi)}{2(1+\epsilon)}}\!.\label{eq9e}
\end{eqnarray}
\end{subequations}
Here we have used the definition
\begin{equation}
\overline{f(\eta,\xi)} = \int_{-\infty}^\infty f(\eta,\xi)\, d\eta. \label{eq10}
\end{equation}

Note that $\hat{E}$ in Eq.~\eqref{eq9e} for $\xi=0$ is exactly the same as $\rho_s(\eta,0)$ of Eq.~\eqref{eq5a}. To find the reduced equation for the slowly varying marginal probability density $P(\xi,\tau;\epsilon)=Q(\xi,\tau;\epsilon)e^{-\frac{\xi^2}{2\epsilon}}$, we use the Chapman-Enskog method \cite{bon10,bon19}:
\begin{subequations} \label{eq11}
\begin{eqnarray}
&&\rho= \hat{E}(\eta,\xi)\, P(\xi,\tau;\epsilon)+ \sum_{j=1}^2\epsilon^j\rho^{(j)}(\eta,\xi;P)+O(\epsilon^3),  \label{eq11a}\\
&& \frac{\partial P}{\partial \tau}= F^{(0)}+ \epsilon F^{(1)}+O(\epsilon^2), \label{eq11b}\\
&&\int_{-\infty}^\infty \rho^{(j)}(\eta,\xi;P)\, d\eta=0, \quad j\geq 1,\label{eq11c}
\end{eqnarray}\end{subequations}
and a similar expansion for $\tilde{\rho}$. Here the $F^{(j)}$ are functionals of $P$ selected so that the hierarchy of linear equations for the $\rho^{(j)}$ (or $\tilde{\rho}^{(j)}$) have bounded solutions.  

Inserting Eqs.~\eqref{eq11} into Eq.~\eqref{eq8a}, we obtain the hierarchy
\begin{subequations}\label{eq12}
\begin{eqnarray}
&&\mathcal{L}\tilde{\rho}^{(0)}=0, \label{eq12a}\\
&&\mathcal{L}\tilde{\rho}^{(1)}=-\mathcal{N}_1\tilde{\rho}^{(0)}+ \frac{\delta\rho^{(0)}}{\delta P}F_0,\label{eq12b}\\
&&\mathcal{L}\tilde{\rho}^{(2)}=-\mathcal{N}_1\tilde{\rho}^{(1)}-\mathcal{N}_2\tilde{\rho}^{(0)} + \frac{\delta\rho^{(1)}}{\delta P}F_0+ \frac{\delta\rho^{(0)}}{\delta P} F_1,     \label{eq12c}
\end{eqnarray}\end{subequations}
etc. 

The solution of the homogeneous equation \eqref{eq12a} is given by Eqs.~\eqref{eq9}. The other equations of the hierarchy are non homogeneous and they have bounded solutions only if the integrals of their right hand side with respect to $\eta\in(-\infty,\infty)$ are zero. The solvability conditions of Eqs.~\eqref{eq12b} and \eqref{eq12c} yield
\begin{subequations}\label{eq13}
\begin{eqnarray}
&& F_0= \int_{-\infty}^\infty\mathcal{N}_1\tilde{\rho}^{(0)}d\eta,\label{eq13a}\\
&& F_1= \int_{-\infty}^\infty[\mathcal{N}_1\tilde{\rho}^{(1)}+\mathcal{N}_2\tilde{\rho}^{(0)}]d\eta,\label{eq13b}
\end{eqnarray}\end{subequations}
respectively, where we have used Eq.~\eqref{eq11c}. Eqs.~\eqref{eq8c}, \eqref{eq9c}, and \eqref{eq13a} yield
\begin{subequations}\label{eq14}
\begin{eqnarray}
&& F_0 =  (\theta_1-\theta_2)\frac{\partial}{\partial\xi}\!\left[P\overline{M(\eta,\xi)}\right]\!,    \label{eq14a}\\
&&M(\eta,\xi)=\frac{\theta_1(\xi-\eta)-\theta_2(\xi+\eta)}{\theta_1\mu_1+\theta_2\mu_2}\mu_1\mu_2\hat{E}\nonumber\\
&&\quad\quad\quad\,\,=\frac{(\theta_1-\theta_2)\xi-2\eta}{\theta_1\mu_1+\theta_2\mu_2}\mu_1\mu_2\hat{E}, \label{eq14b}
\end{eqnarray}
\end{subequations}
where we have used Eq.~\eqref{eq10}. Next, we need to calculate $\rho^{(1)}$ to get $F_1$. Its explicit form is derived in Appendix \ref{ap:b}. Substituting $F_0$ and $F_1$ into Eq.~\eqref{eq11b}, we obtain the sought reduced equation for the marginal probability density $P(\xi,\tau;\epsilon)$:
\begin{widetext}
\begin{subequations}\label{eq15}
\begin{eqnarray}
&&\frac{\partial P}{\partial\tau}= \frac{\partial}{\partial\xi}\!\left[ \mathcal{A}\left(\xi P+\epsilon\frac{\partial P}{\partial\xi}\right) + \mathcal{B}P+ \epsilon \mathcal{D}\frac{\partial P}{\partial\xi}\right]\!,   \label{eq15a}\\
&&\mathcal{A}= 4\theta_1\theta_2\overline{\frac{\mu_1\mu_2\hat{E}}{\theta_1\mu_1\!+\!\theta_2\mu_2}} - \frac{\theta_1\!-\!\theta_2}{1+\epsilon}\overline{\frac{M}{\theta_1\mu_1\!+\!\theta_2\mu_2} \!\left[\int_0^\eta\!\frac{\theta_1\mu_1\!-\!\theta_2\mu_2}{\theta_1\mu_1\!+\!\theta_2\mu_2} d\eta\!-\!\overline{\hat{E}\int_0^\eta\! \frac{\theta_1\mu_1\!-\!\theta_2\mu_2}{\theta_1\mu_1\!+\!\theta_2\mu_2} d\eta }\right]\!},\quad\label{eq15b}\\ 
&&\mathcal{B}=(\theta_1\!-\!\theta_2)\,\overline{M}+\epsilon\, 4\theta_1\theta_2\overline{\frac{\mu_1\mu_2}{\theta_1\mu_1\!+\!\theta_2\mu_2}\frac{\partial\hat{E}}{\partial\xi} } +\epsilon\frac{\theta_1\!-\!\theta_2}{1\!+\!\epsilon}\left[\overline{\frac{(\theta_1\mu_1\!-\!\theta_2\mu_2)\Psi}{\theta_1\mu_1\!+\!\theta_2\mu_2} }+
\overline{ \frac{M\,\tilde{N}}{\theta_1\mu_1+\theta_2\mu_2}}\right]\!, \label{eq15c}\\
&&\tilde{N}=\int_0^\eta\!  \frac{\frac{\theta_1\!-\!\theta_2}{1+\epsilon}\Psi\!-\!(\theta_1\mu_1\!-\!\theta_2\mu_2)\frac{\partial\hat{E}}{\partial\xi}}{(\theta_1\mu_1\!+\!\theta_2\mu_2)\hat{E}}d\eta\!-\!\overline{\hat{E} \!\int_0^\eta\!  \frac{\frac{\theta_1\!-\!\theta_2}{1+\epsilon}\Psi\!-\!(\theta_1\mu_1\!-\!\theta_2\mu_2) \frac{\partial\hat{E}}{\partial\xi}}{(\theta_1\mu_1+\theta_2\mu_2)\hat{E}}d\eta},\label{eq15d}\\
&&\mathcal{D}\!=\!\frac{\theta_1\!-\!\theta_2}{1+\epsilon} \!\left\{ \overline{\frac{\theta_1\mu_1\!-\!\theta_2\mu_2}{\theta_1\mu_1\!\!+\!\theta_2\mu_2}\Phi}\!+\!\overline{\frac{M}{\theta_1\mu_1\!\!+\!\theta_2\mu_2}\!\!\int_0^\eta\!\! \frac{\frac{(\theta_1-\theta_2)\Phi}{(1+\epsilon)\hat{E}}d\eta }{\theta_1\mu_1\!\!+\!\theta_2\mu_2} }
\!-\!\overline{\frac{M}{\theta_1\mu_1\!+\!\theta_2\mu_2}}\,\overline{\hat{E}\! \!\int_0^\eta\!\! \frac{\frac{(\theta_1-\theta_2)\Phi}{(1+\epsilon)\hat{E}}d\eta }{\theta_1\mu_1\!\!+\!\theta_2\mu_2 } }\right\}\!,\quad\label{eq15e}\\
&&\Phi(\eta,\xi)\!=\!\overline{M}(\xi)\!\!\int\!\hat{E}(\eta,\xi)d\eta\!-\!\!\int \!M(\eta,\xi)d\eta,\,\Psi(\eta,\xi)\!=\!\frac{d\overline{M}}{d\xi}\!\!\int\!\hat{E}(\eta,\xi)d\eta\!-\!\frac{\partial}{\partial\xi}\!\int \!M(\eta,\xi) d\eta.\quad\quad\label{eq15f}
\end{eqnarray}\end{subequations}
Eq.~\eqref{eq15a} makes it clear that its stationary solution decaying to zero at $\xi\to\pm\infty$ is proportional to $e^{-\frac{\xi^2}{2\epsilon}}$ if $\theta_1=\theta_2=1$: $\hat{E}$ does not depend on $\xi$ and, according to Eqs.~\eqref{eq15b}-\eqref{eq15e}, we have $\mathcal{A}>0$, $\mathcal{B}=\mathcal{D}=0$. The precise meanings of the indefinite integrals in Eqs.~\eqref{eq15f} depend on the definition of the mobility. They have to be selected so as  to render convergent all integrals in Eqs.~\eqref{eq15}. Defining $\Phi$ and $\Psi$ as
\begin{subequations}\label{eq16}
\begin{eqnarray}
&&\Phi(\eta,\xi)= \frac{1}{2}\!\left[ \int_{-\infty}^\eta\!(\overline{M}\hat{E}- M)\,d\eta+\!\int_\eta^\infty\!(M-\overline{M}\hat{E})\,d\eta\right]\!,  \label{eq16a}\\
&&\Psi(\eta,\xi)=\frac{1}{2}\!\left[\! \int_{-\infty}^\eta\!\!\left(\frac{d\overline{M}}{d\xi}\hat{E}- \frac{\partial M}{\partial\xi}\right)d\eta+\!\int_\eta^\infty\!\!\left(\frac{\partial M}{\partial\xi}-\frac{d\overline{M}}{d\xi}\hat{E}\right) d\eta\right]\!,   \label{eq16b}
\end{eqnarray}\end{subequations}
\end{widetext}
it is clear that $\Phi$ and $\Psi$ vanish as $\eta\to\pm\infty$. With other definitions, it is possible for some integrals appearing in the coefficients $\mathcal{A}$, $\mathcal{B}$ and $\mathcal{D}$ to be divergent. For the mobility of Eq.~\eqref{eq3e}, we show in Appendices \ref{ap:b} and \ref{ap:c} that the singular terms are suppressed by selecting an appropriate current density $J(\xi)$ in the calculation of $\rho^{(1)}$. With the definitions of Eqs.~\eqref{eq16}, there are no divergencies in Eqs.~\eqref{eq15} and we set $J(\xi)=0$. 

\section{Stationary probability density}\label{sec:4}
The overall stationary solution of the FPE is 
\begin{subequations}\label{eq17}
\begin{eqnarray}
&&\rho_s(\eta,\xi;\epsilon)= \hat{E}(\eta,\xi)\hat{F}(\xi), \label{eq17a}\\
&& F(\xi)=e^{-\frac{1}{\epsilon}\int_0^\xi\frac{\mathcal{B}(x)+x\mathcal{A}(x)}{\mathcal{A}(x)+\mathcal{D}(x)}dx},\quad\hat{F}=\frac{F(\xi)}{\int_{-\infty}^\infty F(x)dx}. \quad \label{eq17b}
\end{eqnarray}\end{subequations}
If $\theta_j=1$, $j=1,2$, then $\mathcal{B}=\mathcal{D}=0$, $\mathcal{A}>0$, and Eq.~\eqref{eq17} becomes the equilibrium probability density.

We can calculate stationary averages and correlations of the charges at the capacitors from
\begin{widetext}
\begin{subequations}\label{eq18}
\begin{eqnarray}
&&\langle\eta\rangle_s=\int_{-\infty}^\infty\int_{-\infty}^\infty\eta\hat{E}(\eta,\xi)\hat{F}(\xi)\, d\eta\, d\xi ,\, \langle\xi\rangle_s=\int_{-\infty}^\infty \xi\hat{F}(\xi)d\xi, \label{eq18a}\\
&&\langle q_j\rangle_s=\frac{C_0V_0}{2}\!\left(\frac{\langle\eta\rangle_s}{1+\epsilon}-(-1)^j\frac{\langle\xi\rangle_s}{\epsilon}\right)\! ,    \label{eq18b}\\
&&\langle (q_j-\langle q_j\rangle_s)^2\rangle_s=\frac{C_0^2V_0^2}{4}\!\left\langle\left(\frac{\eta-\langle\eta\rangle_s}{1+\epsilon}-(-1)^j\frac{\xi-\langle\xi\rangle_s}{\epsilon}\right)^2\right\rangle_s\! ,\quad j=1,2,    \label{eq18c}
\end{eqnarray}\end{subequations}

So far, the shape of the diode conductivity $\mu(u)$ has not been used in the derivation. For the diode mobilities of Eq.~\eqref{eq3e} and $\theta_j\neq 1$, $E(\eta,\xi)$ in \eqref{eq9e} can be evaluated exactly in terms of dilogarithm functions but it is not possible to obtain an exact expression for $\overline{E}(\xi)$ or the marginal probability density. However, it is possible to simplify the coefficients in the reduced FPE by using the specific function \eqref{eq3e} for $\xi\gg w$ in the   limit as $w\to 0+$. 
By ignoring boundary layers of width $O(w)$, we have found 
\begin{subequations}\label{eq19}
\begin{eqnarray}
&&\hat{E}(\eta,\xi)\sim\sqrt{\frac{2}{(1\!+\!\epsilon)\pi}}\frac{\Theta(\xi)\Upsilon_2(\eta,\xi) + \Theta(-\xi)\Upsilon_1(\eta,\xi) e^{-(\theta_1\!-\!\theta_2)\frac{(1\!-\theta_1)^2\xi^2}{2\theta_1\theta_2(1\!+\!\epsilon)}} }{\Theta(\xi)\zeta_2(\xi)+\Theta(-\xi)e^{-(\theta_1\!-\!\theta_2)\frac{(1\!-\theta_1)^2\xi^2}{2\theta_1\theta_2(1\!+\!\epsilon)}}\zeta_1(\xi)},  \label{eq19a}\\
&&\Upsilon_j(\eta,\xi)=\Theta(-\eta\!-\!|\xi|)\, e^{-\frac{(1\!-\!\theta_j)(2\!-\!\theta_j)^2\xi^2}{2\theta_j(1\!+\!\epsilon)}-\frac{(\eta-\frac{\theta_1\!-\!\theta_2}{2}\xi)^2}{2(1\!+\!\epsilon)}}\!+\Theta(\eta\!+\!|\xi|)\, e^{-\frac{(\eta-\frac{\theta_1\!-\!\theta_2}{2}\xi)^2}{2(1\!+\!\epsilon)\theta_j}},  \label{eq19b}\\
&&\zeta_j(\xi)= e^{\!-\frac{(1\!-\!\theta_j)(2\!-\!\theta_j)^2\!\xi^2}{2\theta_j(1\!+\!\epsilon)}}\mbox{erfc}\!\left(\!\frac{(2\!-\!\theta_j)|\xi|}{\sqrt{2(1\!+\!\epsilon)}}\!\right)\!+\!\sqrt{\theta_j}\,\mbox{erfc}\!\left(\!\!-\frac{(2\!-\!\theta_j)|\xi|}{\sqrt{2(1\!+\!\epsilon)\theta_j}}\!\right)\!,  \label{eq19c}
\end{eqnarray}
\end{subequations} 
where $\Theta(x)=1$ for $x>0$ and zero otherwise is the unit step function. See Eqs.~\eqref{eqc2} and \eqref{eqc3} of Appendix \ref{ap:c}.  

\begin{center}
\begin{figure}[ht]
\begin{center}
\includegraphics[clip,width=7.5cm]{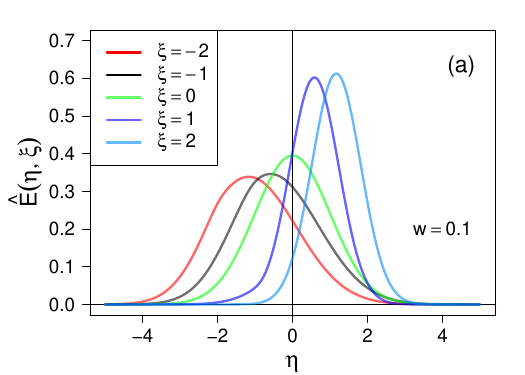}
\includegraphics[clip,width=7.5cm]{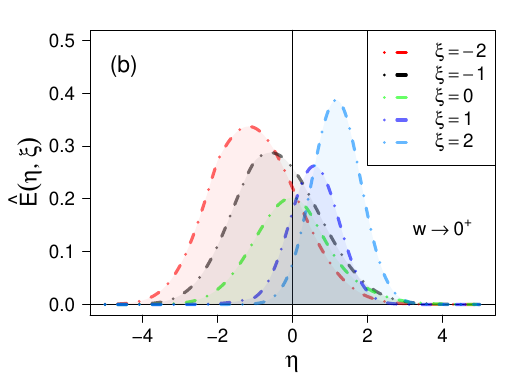}
\includegraphics[clip,width=7.5cm]{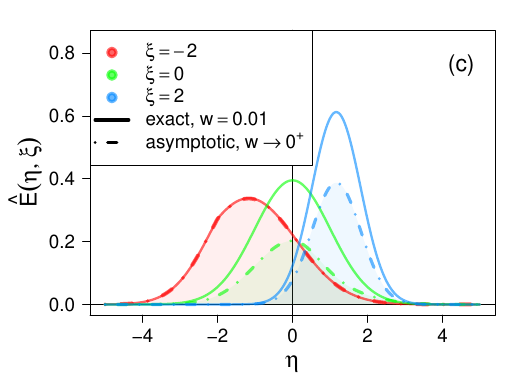}\\
\end{center}
\caption{Function $\hat{E}(\eta,\xi)$ for $\xi=-2,-1,0,1,2$, $\epsilon=0.02$, $T_1= 293$K and $T_2=77$K as {\bf (a)} numerically evaluated for $w=0.1$, {\bf (b)} given by Eq.~\eqref{eq19a} in the limit as $w\to 0+$. {\bf (c)} Comparison between exact and approximate prefactor for $\xi=-2,0,2$ and $w=0.01$.  Note that they coincide for $\xi=-2$ but still differ for $\xi=2$.  \label{fig2}}
\end{figure}
\end{center}
\end{widetext}

\begin{center}
\begin{figure}[ht]
\begin{center}
\includegraphics[clip,width=7.5cm]{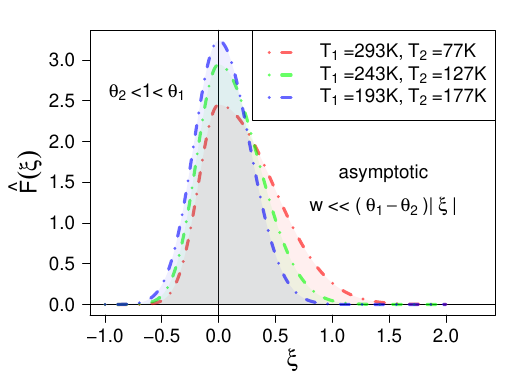}\\
\end{center}
\caption{Approximate stationary marginal probability density for $\epsilon=0.02$, and different temperature differences showing the departure of a symmetric configuration with increasing $(1-\theta_2)$, which is largest for $T_1= 293$K and $T_2=77$K.   \label{fig3}}
\end{figure}
\end{center}

Fig.~\ref{fig2} depicts the prefactor function $\hat{E}(\eta,\xi)$ for $\xi=-2,-1,0,1,2$, $\epsilon=0.02$, $T_1= 293$K and $T_2=77$K. We observe that the numerically calculated prefactor $\hat{E}(\eta,\xi)$ for $w=0.1$ and the approximation \eqref{eq19a} are qualitatively similar for nonzero $|\xi|>w$. There are appreciable quantitative differences for $\xi=2$ and $\xi=1$ but not for $\xi=-2$, $\xi=-1$ and $\xi=0$. Having ignored boundary layers, Eqs.~\eqref{eq19} fail to approximate $\hat{E}(\eta,\xi)$ for very small $|\xi|=O(w)$, as shown by the central curves in Figs.~\ref{fig2}(a) and \ref{fig2}(b). The approximate prefactor is closer to the exact one as $w$ decreases, as shown in Fig.~\ref{fig2}(c).

Now we calculate the stationary marginal probability density of Eq.~\eqref{eq17} for $\theta_2<1<\theta_1$ using the approximate functions in Appendix \ref{ap:c}  that hold for $(\theta_1-\theta_2)|\xi|\gg w$. We find
\begin{subequations}\label{eq20}
\begin{eqnarray}
\hat{F}(\xi)=\frac{\sqrt{\frac{2}{\pi\epsilon}}}{\sum_{j=1}^2\sqrt{\frac{7}{\theta_j}-5+\theta_j}}\sum_{j=1}^2\Theta_j\, e^{-\frac{\theta_j\xi^2}{2\epsilon(7-5\theta_j +\theta_j^2)}}, \quad\quad\label{eq20a}
\end{eqnarray}
with $\Theta_j=\Theta((-1)^j\xi)$, which is shown in Figure \ref{fig3}. In this equation, $7-5\theta_j +\theta_j^2=1+(2-\theta_j)+(2-\theta_j)^2\in(1,7)$ because $0<\theta_j<2$. Note the rapid decay of the density with $|\xi|$ and its asymmetric shape that assigns higher probability for the capacitor at lower temperature ($\xi>0$) to gather more charge than the capacitor at higher temperature. For equal temperatures, the exponential factors in Eq.~\eqref{eq20a} are $e^{-\frac{\xi^2}{6\epsilon}}$ instead of $e^{-\frac{\xi^2}{2\epsilon}}$ for the equilibrium distribution. These exponentials are different because the nonequilibrium distribution has been obtained assuming that $(\theta_1-\theta_2)|\xi|\gg 1$, which excludes the equal temperature case.

\begin{widetext}
We now use Eqs.~\eqref{eq19} and \eqref{eq20a} and approximate integrals by the Laplace method \cite{bender}, thereby obtaining the stationary averages
\begin{eqnarray}
&&\langle\eta\rangle_s\sim\frac{\frac{\theta_1-\theta_2}{\sqrt{2\pi}}}{\sqrt{\frac{7}{\theta_1}-5+\theta_1}+\sqrt{\frac{7}{\theta_2}-5+\theta_2} }\! \left[ \sqrt{1+\epsilon}\!\left(\frac{\sqrt{\frac{7}{\theta_1}-5+\theta_1}}{1+\sqrt{\theta_1}} -\frac{\sqrt{\frac{7}{\theta_2}-5+\theta_2}}{1+\sqrt{\theta_2}} \right) 
+\sqrt{\epsilon}(\theta_1-\theta_2)\!\left(\frac{7}{\theta_1\theta_2}-1\right)\!\right]\!,\quad \label{eq20b}\\
&&\langle\xi\rangle_s\sim \sqrt{\frac{2\epsilon}{\pi}}\frac{\theta_1-\theta_2}{\sqrt{\frac{7}{\theta_1}-5+\theta_1}+\sqrt{\frac{7}{\theta_2}-5+\theta_2} }\! \left(\frac{7}{\theta_1\theta_2}-1\right)\!,\label{eq20c}\\
&&\langle\eta^2\rangle_s\sim1+\epsilon +\frac{(1\!+\!\epsilon) (\theta_1\!-\!\theta_2)}{\!\sum_j\sqrt{\frac{7}{\theta_j}\!-\!5\!+\!\theta_j}}\!\sum_{j=1}^2\! \frac{(-1)^j}{1\!+\!\sqrt{\theta_j}}\left[\frac{\sqrt{\epsilon}(2\!-\!\theta_j)}{\pi\sqrt{1\!+\!\epsilon}}\!\left(\frac{7}{\theta_j}\!-\!5\!+\!\theta_j\right)\!-\frac{\sqrt{7\!-\!5\theta_j\!+\!\theta_j^2}}{2}\right]\! 
+ \!\left(\!\frac{\theta_1\!-\!\theta_2}{2}\!\right)^2\!\langle\xi^2\rangle_s,    \label{eq20d}\\
&&\langle\xi\eta\rangle_s\sim\frac{\theta_1-\theta_2}{2}\!\left[\langle\xi^2\rangle_s+\frac{2\sqrt{\epsilon(1+\epsilon)}}{\pi\sum_{j=1}^2\sqrt{\frac{7}{\theta_j}-5+\theta_j}}\sum_{j=1}^2\frac{\frac{7}{\theta_j}-5+\theta_j}{1+\sqrt{\theta_j}}\right]\!, \label{eq20e}\\
&&\langle\xi^2\rangle_s\sim\epsilon\frac{\left(\frac{7}{\theta_1}-5+\theta_1\right)^\frac{3}{2}+\left(\frac{7}{\theta_2}-5+\theta_2\right)^\frac{3}{2}}{\sqrt{\frac{7}{\theta_1}-5+\theta_1}+\sqrt{\frac{7}{\theta_2}-5+\theta_2} }=\epsilon\!\left[\frac{14}{\theta_1\theta_2}-8-\sqrt{\prod_{j=1}^2\!\left(\frac{7}{\theta_j}-5+\theta_j\right)\!}\,\right]\!. \label{eq20f}
\end{eqnarray}
\end{widetext}
\end{subequations}

\section{Evolution of the marginal density}\label{sec:5}
The reduced FPE for the marginal probability density $P(\xi,\tau)$ has been derived in the limit as $\epsilon\to 0+$. By using the specific function \eqref{eq3e} for $\xi\gg w$ in the limit as $w\to 0+$, we can study the transient stage using the reduced FPE \eqref{eq15a}, which can be written as
\begin{subequations}\label{eq21}
\begin{eqnarray}
&&\frac{\partial g}{\partial \tau}+\!\left(\xi\mathcal{A}+\mathcal{B}\right)\frac{\partial g}{\partial\xi}=\epsilon  \frac{\partial}{\partial\xi}\!\left[(\mathcal{A}+\mathcal{D})\frac{\partial g}{\partial\xi}\right]\!, \label{eq21a}\\
&& P(\xi,\tau)=F(\xi)\, g(\xi,\tau).   \label{eq21b}
\end{eqnarray}\end{subequations} 
Ignoring $O(\epsilon)$ terms, we get the wave front solution
\begin{subequations}\label{eq22}
\begin{eqnarray}
&&g\propto\Theta(\Xi(\tau)-\xi),\quad\mbox{ where}\label{eq22a}\\
&&\frac{d\Xi }{d\tau}= \Xi\,\mathcal{A}(\Xi)+\mathcal{B}(\Xi). \label{eq22b}
\end{eqnarray}\end{subequations}
Eq.~\eqref{eq22a} follows from an initial condition having $P=0$ for $\xi>\Xi(0)>0$ and $P=\hat{F}(\xi)$ behind the front. If $\theta_1=\theta_2=1$ this corresponds to the equilibrium density invading a forbidden region of charge difference ahead of the front. For a front advancing to the left, $g\propto\Theta(\xi-\Xi(\tau))$, $\Xi(\tau)<0$. 

\subsection{Equal temperature at diodes}
\begin{center}
\begin{figure}[ht]
\begin{center}
\includegraphics[clip,width=7.5cm]{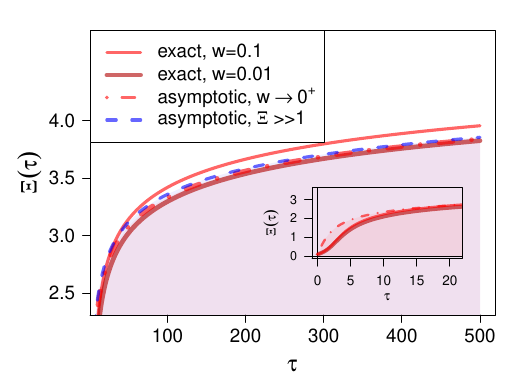}
\end{center}
\caption{Rapid slowing down of the advancing front  $\Xi(\tau)$ given by Eqs.~\eqref{eq23a} with $w=0.01, 0.1$, by the approximation \eqref{eq23b} ($w\to 0$) and by Eq.~\eqref{eq25} ($\Xi\gg 1$) for $\tau>20$. The initial condition is $\Xi (0)= 0.1\geq w$.  Inset: evolution for $0<\tau<20$. \label{fig4}}
\end{figure}
\end{center}

Let us first consider the case $\theta_j=1$ studied in \cite{thi23}. Then Eq.~\eqref{eq22b} becomes
\begin{subequations}\label{eq23}
\begin{eqnarray}
&&\frac{d\Xi}{d\tau}\!=\!\Xi\,\mathcal{A}(\Xi)\!=\!\frac{2\,\Xi}{\sqrt{2\pi(1\!+\!\epsilon)}}\!\int_{-\infty}^\infty\! \frac{e^{-\frac{\eta^2}{2(1+\epsilon)}}d\eta}{1+e^{\eta/w}\cosh\frac{\Xi}{w}},\quad \label{eq23a}\\
&&\frac{d\Xi}{d\tau}\sim \sqrt{\frac{2(1+\epsilon)}{\pi}}\,  e^{-\frac{\Xi^2}{2(1+\epsilon)}}\,\mbox{sign}\,\Xi,\quad\mbox{as $w\to 0+$.} \label{eq23b}
\end{eqnarray}
\end{subequations}
This is similar to the case of small temperature considered in Appendix B of \cite{thi23} for a single diode. Adapting it to the present case, the velocity of characteristics is exponentially small as $\Xi\to\infty$, so the front slows down dramatically as it advances. Figure \ref{fig4} shows that the front advances rapidly from very low charge differences $\Xi(0)=0.1$ to values about between 2 and 3 and then it increases extremely slowly. The approximation \eqref{eq23b} for $w\ll 1$ captures the long time behavior of the solution of Eq.~\eqref{eq23a}. 

In practice, the front becomes frozen at a finite value of $\xi$. It is possible to estimate approximately the time it takes the front to advance from $\xi = \Xi>0$ to $\xi = \Xi+ \delta\Xi$, $0<\delta\Xi\ll\Xi$, 
\begin{eqnarray}
\delta \tau= \frac{e^{\frac{\Xi^2}{2(1+\epsilon)}}\sqrt{\pi}}{\sqrt{2(1+\epsilon)}} \,\delta\Xi. \label{eq24}
\end{eqnarray}
This time becomes exponentially large as $\Xi\gg 1$. The decay of the ensemble-averaged charge slows down due to the slowing of the front. As $\tau\to\infty$, the solution of Eq.~\eqref{eq23} can be obtained from the asymptotic expansion of $x e^{-x^2}=\delta$ as $\delta\to 0$, which gives 
\begin{eqnarray}
\Xi(\tau)\sim \!\sqrt{2(1+\epsilon)}\!\left[\!\sqrt{\ln(q\tau)}+\frac{1}{4}\frac{\ln\ln(q\tau)}{\sqrt{\ln(q\tau)}}\right]\!, \, q=\frac{2}{\!\sqrt{\pi}}, \quad\label{eq25}
\end{eqnarray}
as explained in Appendix \ref{ap:e}. Fig.~\ref{fig4} also compares the solution of Eqs.~\eqref{eq23} with Eq.~\eqref{eq25}, which captures quite well the trend of the solution.

If we approximate
\begin{eqnarray}
P= \frac{\Theta(\Xi(\tau)-\xi)}{\int_{-\infty}^\Xi e^{-\xi^2/(2\epsilon)}d\xi}\, e^{-\xi^2/(2\epsilon)}, \label{eq26}
\end{eqnarray}
then the uniform value of $g$ behind the front is not exactly time-independent, consistent with the advection equation Eq.~\eqref{eq21}. This is an error associated with the step function approximation to $g$. The error in normalization is exponentially small for $\Xi \gg \sqrt{\epsilon}$, and has negligible effect on the estimate of ensemble-averaged charge, which is now
\begin{eqnarray}
\langle \xi\rangle\!=\! \frac{\int_{-\infty}^\Xi\! \xi\, e^{-\frac{\xi^2}{2\epsilon}}d\xi}{\sqrt{2\pi\epsilon}}\!  =\!- \sqrt{\frac{\epsilon}{2\pi}} e^{-\frac{\Xi^2}{2\epsilon}}. \label{eq27}
\end{eqnarray}
Note that for $\Xi(0)=0$, the capacitors have nonzero equal and opposite initial charges $\langle q_j\rangle= \frac{(-1)^jC_0V_0}{\sqrt{2\pi\epsilon}}$ and $\Xi(\tau)=0$. For an infinitesimal $\Xi(0)$, the relative increase of average charge difference over the time in which the front advances from $\xi = \Xi$ to $\xi = \Xi + \delta\Xi$ is
\begin{subequations}\label{eq28}
\begin{eqnarray}
\frac{\delta\langle \xi\rangle}{\langle \xi\rangle} \sim \frac{\Xi\,\delta \Xi}{\epsilon}. \label{eq28a}
\end{eqnarray}
From Eqs.~\eqref{eq24} and \eqref{eq28a}, the time required for a given relative increase of charge difference, $a=\delta\langle \xi\rangle/\langle \xi\rangle$, is
\begin{eqnarray}
\delta\tau \sim \epsilon \frac{e^\frac{\Xi^2}{2(1+\epsilon)}}{\Xi\sqrt{2(1+\epsilon)}} \, a\sqrt{\pi}. \label{eq28b}
\end{eqnarray}\end{subequations}
This time becomes exponentially large as $\Xi\to\infty$.

The diffusion in Eq.~\eqref{eq21} smooths out the front at $\xi = \Xi(\tau)$. To see how this works, we examine the equation for the gradient of $g$ near the front, $r=\partial g/\partial\xi$. Eq.~\eqref{eq21} becomes
\begin{eqnarray}
\frac{\partial r}{\partial \tau}+\frac{\partial}{\partial\xi}\left(\xi\mathcal{A}r -\epsilon\frac{\partial(\mathcal{A} r)}{\partial\xi}\right)\!=0. \label{eq29}
\end{eqnarray}
This is a diffusion-convection equation (different from the FPE) and therefore $r$ is locally conserved. Hence, the total change in $g$ across the front is conserved. We represent $r$ in ``traveling wave'' form,
\begin{equation}
r=\mathcal{R}(\zeta,\tau), \quad \zeta=\frac{\xi-\Xi(\tau)}{\sqrt{\epsilon}}. \label{eq30}
\end{equation}
Eqs~\eqref{eq29} and \eqref{eq30} provide $\mathcal{R}$ as a solution of
\begin{widetext}
\begin{eqnarray*}
\frac{\partial\mathcal{R}}{\partial\tau}+\frac{1}{\sqrt{\epsilon}}\frac{\partial}{\partial\zeta}\!\left\{\!\left[ (\Xi+\sqrt{\epsilon}\zeta)\mathcal{A}(\Xi+\sqrt{\epsilon}\zeta)-\dot{\Xi}\right]\!\mathcal{R} -\sqrt{\epsilon}\frac{\partial}{\partial\zeta}\!\left[\mathcal{A}(\Xi+\sqrt{\epsilon}\zeta)\mathcal{R}\right]\right\}\!=0,
\end{eqnarray*}
(where $\dot{\Xi}=d\Xi/d\tau$) or, evoking Eq.~\eqref{eq22b} for $\dot{\Xi}$,
\begin{eqnarray*}
\frac{\partial\mathcal{R}}{\partial\tau}+\frac{\partial}{\partial\zeta}\!\left\{\!\left[ \frac{(\Xi+\sqrt{\epsilon}\zeta)\mathcal{A}(\Xi+\sqrt{\epsilon}\zeta)-\Xi\mathcal{A}(\Xi)}{\sqrt{\epsilon}}\right]\!\mathcal{R} -\frac{\partial}{\partial\zeta}\!\left[\mathcal{A}(\Xi+\sqrt{\epsilon}\zeta)\mathcal{R}\right]\right\}\!=0. 
\end{eqnarray*}
\end{widetext}
In the limit as $\epsilon\to 0$,
\begin{eqnarray}
\frac{\partial\mathcal{R}}{\partial\tau}+ \frac{\partial}{\partial\zeta}\!\left\{(\Xi\mathcal{A})'(\Xi) \zeta\mathcal{R} - \frac{\partial}{\partial\zeta}[ \mathcal{A}(\Xi)\,\mathcal{R}]\right\}\!=0. \label{eq31}
\end{eqnarray}
This equation has a Gaussian as solution, 
\begin{subequations}\label{eq32}
\begin{eqnarray}
\mathcal{R}=\frac{1}{\sqrt{2\pi\sigma}}\, e^{-\zeta^2/(2\sigma)},\label{eq32a}
\end{eqnarray}
 whose variance satisfies
\begin{eqnarray}
\dot{\sigma}-2(\Xi\mathcal{A})'(\Xi)\,\sigma = 2\mathcal{A}(\Xi). \label{eq32b}
\end{eqnarray}
From Eq.~\eqref{eq23}, this equation becomes
\begin{eqnarray}
\frac{d\sigma}{d\Xi}-\frac{2(\Xi\mathcal{A})'(\Xi)}{\Xi\mathcal{A}(\Xi)}\,\sigma = \frac{2}{\Xi}, \label{eq32c}
\end{eqnarray}
so that 
\begin{eqnarray}
\sigma(\Xi)=  2[\Xi\mathcal{A}(\Xi)]^2\int_{\xi_0}^\Xi\frac{d\xi}{\xi^3[\mathcal{A}(\xi)]^2}. \label{eq32d}
\end{eqnarray}
Note that the right hand side of Eq.~\eqref{eq23} is odd in $\Xi$ and therefore $\Xi(\tau)$ with $\Xi(0)=\Xi_0>0$ produces the solution $-\Xi(\tau)$ for the same equation with initial condition $\Xi(0)=-\Xi_0$. Similarly, $\sigma(\Xi)$ is even in $\Xi$ provided $\xi_0=0$. 

\begin{center}
\begin{figure}[ht]
\begin{center}
\includegraphics[clip,width=7.5cm]{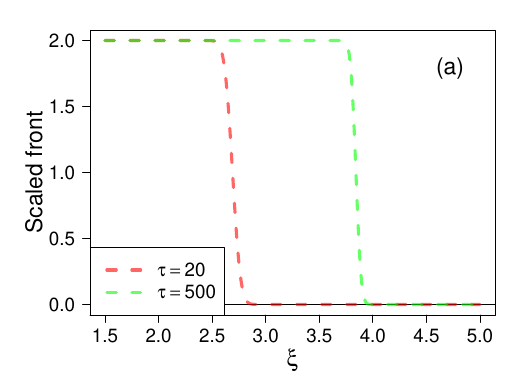}
\includegraphics[clip,width=7.5cm]{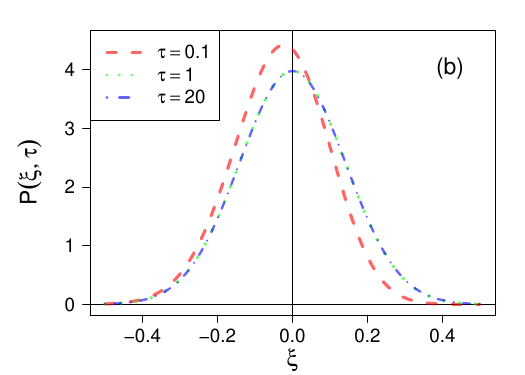}
\end{center}
\caption{{\bf (a)} Scaled marginal probability density  erfc$[(\xi-\Xi(\tau))/\sqrt{2\epsilon\sigma(\tau)}]$ at times 20 and 500 for $\Xi(0)=0.1$. {\bf (b)} Marginal probability density at times $t=0.1, 1, 20$. The curves at times $t=1,20$ are indistinguishable. \label{fig5}}
\end{figure}
\end{center}

Suppose the front has advanced far enough so $\Xi\gg 1$. Then using $\mathcal{A}(\Xi)$ from Eq.~\eqref{eq23}, Eq.~\eqref{eq32c} reduces to
\begin{eqnarray}
\frac{d\sigma}{d\Xi}+\frac{2\,\Xi}{1+\epsilon}\sigma = \frac{2}{\Xi}. \label{eq32e}
\end{eqnarray}
As $\Xi\to\infty$, there is an asymptotic solution 
\begin{eqnarray}
\sigma\sim \sum_{n=1}^\infty(n-1)!\left(\frac{\sqrt{1+\epsilon}}{\Xi}\right)^{2n-1},  \label{eq32f}
\end{eqnarray}
\end{subequations}
and the front thickness narrows as it propagates further to the right. Integrating $r$ over $\xi$,  we get a better approximation to the marginal probability density:
\begin{subequations}\label{eq33}
\begin{eqnarray}
&&P(\xi,\tau)=  \frac{1}{Z(\tau)} e^{-\frac{\xi^2}{2\epsilon}}\,\mbox{erfc}\!\left(\frac{|\xi-\Xi(\tau)|}{\sqrt{2\epsilon\sigma(\tau)}}\right)\!,  \label{eq33a}\\
&&Z(\tau)=\int_{-\infty}^\infty e^{-\frac{\xi^2}{2\epsilon}}\,\mbox{erfc}\!\left(\frac{|\xi-\Xi(\tau)|}{\sqrt{2\epsilon\sigma(\tau)}}\right) d\xi. \label{eq33b}
\end{eqnarray}
\end{subequations}
Fig.~\ref{fig5}(a) shows the moving error function of Eq.~\eqref{eq33a} for two different times and initial conditions outside the boundary layer. That the front thickness narrows as it propagates is clearly illustrated. However, the marginal probability density is essentially the marginal equilibrium density at these two times due to the sharp prefactor $e^{-\xi^2/(2\epsilon)}$ as further illustrated by the marginal probability density at times 0.1, 1, 20 depicted in Fig.~\ref{fig5}(b)

\subsection{Different temperatures at diodes}
The exponential slowing down that appears for diodes at the same temperature is compensated by the terms containing $M/(\theta_1\mu_1+\theta_2\mu_2)$ in Eqs.~\eqref{eq15} that are $O(\xi)$ as $\xi\to\infty$ and get multiplied by a nonzero factor if $\theta_1\neq\theta_2$. The marginal probability density corresponding to the front is
\begin{subequations}\label{eq34}
\begin{eqnarray}
&&P(\xi,\tau)\!=\!  \frac{1}{Z(\tau)} e^{-\frac{1}{\epsilon}\int_0^\xi\frac{\mathcal{B}(x)+x\mathcal{A}(x)}{\mathcal{A}(x)+\mathcal{D}(x)}dx}\,\mbox{erfc}\!\left(\!\frac{\xi-\Xi(\tau)}{\sqrt{2\epsilon\sigma(\tau)}}\!\right)\!, \quad \label{eq34a}\\
&&Z(\tau)\!=\!\!\int_{-\infty}^\infty\! e^{-\frac{1}{\epsilon}\int_0^\xi\frac{\mathcal{B}(x)+x\mathcal{A}(x)}{\mathcal{A}(x)+\mathcal{D}(x)}dx}\,\mbox{erfc}\!\left(\!\frac{\xi-\Xi(\tau)}{\sqrt{2\epsilon\sigma(\tau)}}\!\right)d\xi, \label{eq34b}
\end{eqnarray}
\end{subequations}
where the front evolves according to Eq.~\eqref{eq22b}.

The extra terms for the case of different temperatures provide a shorter relaxation stage towards the stationary solution  \eqref{eq17}. In fact, setting $r=\partial g/\partial\xi$ in Eqs.~\eqref{eq21} - \eqref{eq22b} for $\theta_j\neq 1$, $j=1,2$, replaces Eq.~\eqref{eq29} by
\begin{eqnarray}
\frac{\partial r}{\partial\tau}+\frac{\partial}{\partial\xi}\left[(\xi\mathcal{A}+\mathcal{B})\, r -\epsilon\frac{\partial[(\mathcal{A}+\mathcal{D}) r]}{\partial\xi}\right]\!=0. \label{eq35}
\end{eqnarray}
Using now Eqs.~\eqref{eq30} and \eqref{eq32a}, we get
\begin{subequations}\label{eq36}
\begin{eqnarray}
&&\frac{d\Xi}{d\tau}=\Xi\,\mathcal{A}(\Xi)+\mathcal{B})(\Xi),   \label{eq36a}\\
&&\frac{d\sigma}{d\tau}-2(\Xi\mathcal{A}\!+\!\mathcal{B})'(\Xi)\,\sigma = 2[\mathcal{A}(\Xi)\!+\!\mathcal{D}(\Xi)] \Longrightarrow\label{eq36b}\\
&&\frac{d\sigma}{d\Xi}-\frac{2(\Xi\mathcal{A}\!+\!\mathcal{B})'(\Xi)}{\Xi\mathcal{A}(\Xi)\!+\!\mathcal{B}(\Xi)}\,\sigma = 2\,\frac{\mathcal{A}(\Xi)\!+\!\mathcal{D}(\Xi)}{\Xi\mathcal{A}(\Xi)\!+\!\mathcal{B}(\Xi)}, \label{eq36c}
\end{eqnarray}
instead of Eq.~\eqref{eq32c}. The solution of this equation is 
\begin{eqnarray}
\sigma(\Xi)=  2[\Xi\mathcal{A}(\Xi)+\mathcal{B}(\Xi)]^2\int_{\xi_0}^\Xi\frac{[\mathcal{A}(\xi)+\mathcal{D}(\xi)]\, d\xi}{[\xi\mathcal{A}(\xi)+\mathcal{B}(\xi)]^3}.\quad \label{eq36d}
\end{eqnarray}\end{subequations}
For the specific diode mobility of Eq.~\eqref{eq3e} with $w\to 0+$, different temperatures, and $|\xi|\gg 1$, the coefficients in Eqs.~\eqref{eq36} become 
\begin{widetext}
\begin{subequations}\label{eq37}
\begin{eqnarray}
&&\mathcal{A}=\sum_{j=1}^2\Theta_j\mathcal{A}_j, \quad\mathcal{A}_j\sim \sqrt{\frac{(1\!+\!\epsilon)\theta_j}{2\pi}}\!\left(\frac{2}{|\xi|}\!+\!\frac{(1\!-\!\theta_j)|\xi|}{1+\epsilon}\right)\! e^{-\frac{(2-\theta_j)^2\xi^2}{2(1\!+\!\epsilon)\theta_j}} , \label{eq37a}\\
&&\mathcal{D}=\sum_{j=1}^2\Theta_j \mathcal{D}_j, \quad\mathcal{D}_j\!\sim\! \frac{1-\theta_j}{\sqrt{2\pi\theta_j(1\!+\!\epsilon)}} (7\!-\!6\theta_j\!+\!\theta_j^2)|\xi| e^{-\frac{(2\!-\!\theta_j)^2\xi^2}{2(1+\epsilon)\theta_j}}, \quad\quad\label{eq37b}\\
&&\mathcal{B}=\sum_{j=1}^2\Theta_j\mathcal{B}_j, \quad\mathcal{B}_j\sim (\theta_1-\theta_2)\sqrt{\frac{1\!+\!\epsilon}{2\pi\theta_j}} \, e^{-\frac{(2\!-\!\theta_j)^2\xi^2}{2(1\!+\!\epsilon)\theta_j}}\!\left[1-\frac{\epsilon(1\!-\!\theta_j)^2(5-4\theta_j)\xi^2}{4\theta_j(1+\epsilon)^2}\right]\!,  \label{eq37c}\\
&&\frac{\xi\mathcal{A}+\mathcal{B}}{\mathcal{A}+\mathcal{D}} \sim \sum_{j=1}^2\frac{\Theta((-1)^j\xi)\,\theta_j\xi}{7-5\theta_j+\theta_j^2}. \quad \label{eq37d}
\end{eqnarray}
where sign$\xi=(-1)^j$, $\Theta_j=\Theta((-1)^j\xi)$, $j=1,2$; see Appendix \ref{ap:c}. 
\end{subequations} 

\begin{center}
\begin{figure}[ht]
\begin{center}
\includegraphics[clip,width=7.5cm]{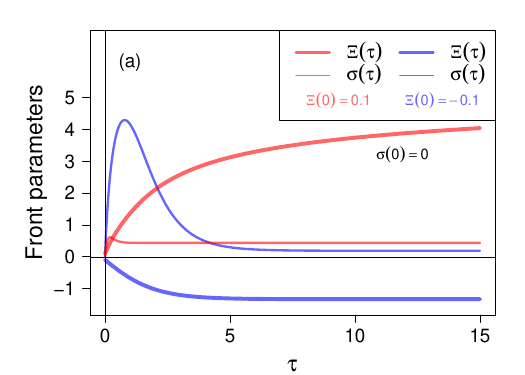}
\includegraphics[clip,width=7.5cm]{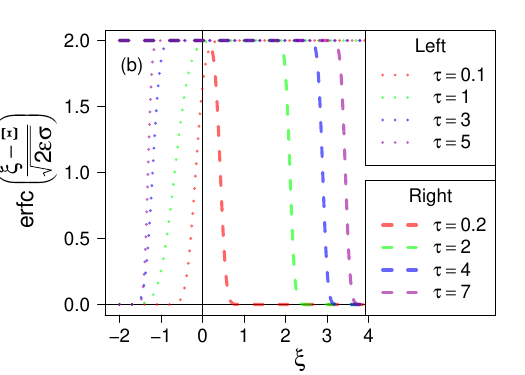}
\end{center}
\caption{{\bf (a)} Position and variance of the marginal probability density front for initial conditions $\Xi(0)=\pm 0.1$, $\sigma(0)=0$. {\bf (b)} Scaled  marginal probability density erfc$[(\xi-\Xi(\tau))/\sqrt{2\epsilon\sigma(\tau)}]$ at the indicated times for left and right moving fronts. Note that the variance of the left moving front first increases and then decreases with time. The nonmonotone region of $\sigma(\tau)$ is much smaller for the right moving front. \label{fig6}}
\end{figure}
\end{center}
 
As $|\xi|\gg 1$, the probability density for different temperatures $\theta_2<1<\theta_1$ is given by Eqs.~\eqref{eq17}, \eqref{eq34} and \eqref{eqc3}. The approximations \eqref{eq37} and \eqref{eq36d} yield
\begin{subequations}\label{eq38}
\begin{eqnarray}
&&\rho(\eta,\xi,\tau)=\hat{E}(\eta,\xi)\,\hat{F}(\xi,\tau), \label{eq38a}\\
&&\hat{F}(\xi,\tau)=\frac{1}{Z(\tau)}\left[e^{-\frac{\theta_1\xi^2/(2\epsilon)}{7-5\theta_1+\theta_1^2}}\mbox{erfc}\!\left(\!\frac{\Xi(\tau)\!-\!\xi}{\sqrt{2\epsilon \sigma(\tau)}}\right)\!\Theta(-\xi)\!+\!e^{-\frac{\theta_2\xi^2/(2\epsilon)}{7-5\theta_2+\theta_2^2}}\mbox{erfc}\!\left(\frac{\xi-\Xi(\tau)}{\!\sqrt{2\epsilon\sigma(\tau)}}\right)\!\Theta(\xi)\right]\!\!,\quad \label{eq38b}\\
&&Z(\tau)=\!\int_0^\infty\!e^{-\frac{\theta_1\xi^2/(2\epsilon)}{7-5\theta_1+\theta_1^2}}\mbox{erfc}\!\left(\!\frac{\xi\!-\!|\Xi(\tau)|}{\sqrt{2\epsilon\sigma(\tau)}}\right)\!d\xi+\!\int_0^\infty\!e^{-\frac{\theta_2\xi^2/(2\epsilon)}{7-5\theta_2+\theta_2^2}}\mbox{erfc}\!\left(\!\frac{\xi\!-\!\Xi(\tau)}{\sqrt{2\epsilon\sigma(\tau)}}\right)\!d\xi,  \label{eq38c}\\
&&\sigma\!\sim\! 4\pi(1\!+\!\epsilon)^2\!\sum_{j=1}^2\!\Theta((-1)^j\Xi) \frac{\theta_j(7\!-\!5\theta_j\!+\!\theta_j^2)}{(2-\theta_j)\,\Xi^2},\,\,\Xi\! \sim\! \frac{\sqrt{2\theta_j(1\!+\!\epsilon)}}{2-\theta_j}\!\left[\!\sqrt{\ln(q\tau)}+\frac{3}{4}\frac{\ln\ln(q\tau)}{\sqrt{\ln(q\tau)}}\!\right]\!, \quad \label{eq38d}
\end{eqnarray}\end{subequations} 
\end{widetext}
with $q=\frac{(\theta_1\!-\!\theta_2)\theta_j}{\sqrt{\pi}(2\!-\!\theta_j)}$ as $\tau\to\infty$; see Appendix \ref{ap:e}. Again the front thickness narrows as it advances but it does so at a faster rate than in the case of equal temperatures. Fig.~\ref{fig6}(a) shows the evolution of the front for negative and positive values of $\xi$ as well as the evolution of its variance obtained from Eqs.~\eqref{eq36a} and \eqref{eq36b}. The variance reaches a local maximum at short times and then tends to a constant value. The error functions in Eq.~\eqref{eq38b} evolve as depicted in Fig.~\ref{fig6}(b). The left moving front freezes more rapidly than the right moving front and its variance, which initially is very large, decays faster.

\begin{center}
\begin{figure}[ht]
\begin{center}
\includegraphics[clip,width=7.5cm]{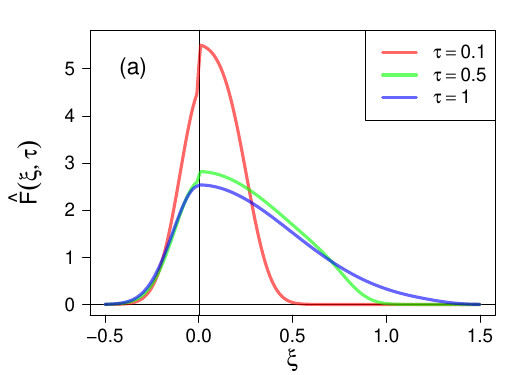}
\includegraphics[clip,width=7.5cm]{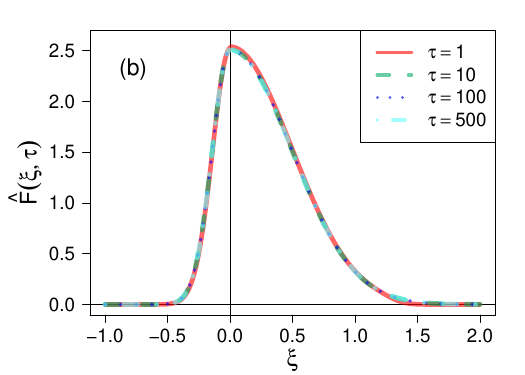}
\end{center}
\caption{Evolution of  the marginal probability density $\hat{F}(\xi,\tau)$ at {\bf (a)} short times $\tau=0.1,0.5,1$ and {\bf (b)} long times $\tau=1,10,100,500$. \label{fig7}}
\end{figure}
\end{center}

 The stationary probability density of Eq.~\eqref{eq20a} is reached as $|\Xi|\to\infty$ in Eqs.~\eqref{eq38}. Figure \ref{fig7} shows the evolution of the marginal probability density. Ignoring the boundary layer at $|\xi|\sim w$, together with the large initial variance of the left moving wave front, produces a discontinuity at $\xi=0$ which is reduced at short times as depicted in Fig.~\ref{fig7}(a) whereas after $\tau=10$, the marginal probability density has achieved the stationary value of Fig.~\ref{fig3}, as shown in Fig.~\ref{fig7}(b).

\section{Discussion and final remarks}\label{sec:6}
Here we have considered an energy harvesting circuit that stems from the electrical and mechanical rectifiers of thermal fluctuations proposed by Brillouin \cite{bri50} and Feynman \cite{feynman}, respectively. Sixty years ago, different authors studied thermal fluctuations and effects of shot noise that could be rectified using nonlinear electrical diodes \cite{vka60,lan62}. In the late 1990s, Sokolov theoretically studied systems using capacitors and diodes and found that charge is stored on the capacitor in the steady state when there is a temperature difference between the diodes \cite{sok98,sok99}. Extraction of electric energy from thermal effects may be achieved using thermoelectric materials \cite{gol14}, particularly low dimensional materials that have a higher figure of merit \cite{dre99,mao16,nin24}.  

We have studied a rectifying circuit that can be used to charge capacitors out of thermal fluctuations, i.e., for energy harvesting. The circuit consists of two diodes placed in opposition and two storage capacitors coupled to a freestanding graphene sheet through a STM. The freestanding graphene acts as a capacitor of variable capacitance that is much smaller than that of the storage capacitors \cite{thi20}. The small capacitance ratio causes the system to evolve rapidly to a quasi stationary state with a specific average value of the total charge in the circuit, which is zero if the system is in contact with a single thermal bath \cite{thi23,ami25}. On a much longer time scale, the system evolves to a stationary state or to thermal equilibrium depending on whether there are temperature differences or not. During this long transient, the storage capacitors may charge and be used for energy harvesting, steadily in the case of different temperatures or  disconnecting them first in the case of a single temperature. 

The analysis of this energy harvesting system is delicate due to the small capacitance ratio and other small parameters associated to the nonlinear mobilities of the diodes. In the case of a single temperature, the evolution of the probability density should always have the canonical equilibrium density as a stationary solution of the Fokker-Planck equation although the terms of the electrostatic energy due to graphene and the storage capacitors are vastly different: a term dependent on charge differences becomes very large as the capacitance ratio goes to zero. The corresponding factor in the canonical distribution is exponentially small. Here we tackle this problem of exponential asymptotics by extracting the exponentially small term as a prefactor of the evolving probability density. After a {\em very fast initial stage} (cf Appendix \ref{ap:a} and \cite{thi23}), the probability density reaches a quasi stationary state that depends on the nondimensional sum of the charges at the storage capacitors multiplied by a slowly varying marginal probability density that depends on charge differences. Here we use a Chapman-Enskog procedure to describe the long evolution of the marginal probability density from the quasi stationary state to the final stationary state (different temperatures) or to thermal equilibrium (single temperature) \cite{bon19}. The marginal probability density satisfies a Smoluchowski advection-diffusion equation. That the results of this procedure agree well with numerical simulations of the full FPE is shown in the second paper of this series. 

In this paper, we further approximate the equation for the marginal probability that describes the slow evolution towards the final stationary state. To do so, we exploit another small parameter occurring for a specific form of the diodes nonlinear mobility. This extra parameter is a nondimensional diode temperature which is small compared to the average temperature of the diodes. We approximate the expressions appearing in the description of the marginal probability density, which allows us to visualize the evolution towards the final stationary state. Except for boundary layers, the quasi stationary state consists of different Gaussian exponentials that hold in different regions of the charge space. The stationary marginal probability density is either the equilibrium (single temperature) or the sum of two Gaussian functions that are different for positive and negative charge differences (different temperatures). In the quasistationary state after the initial stage, the storage capacitors have equal and opposite average charges in the case of a single temperature (see Appendix \ref{ap:a}). These charges can be extracted if the capacitors are disconnected after the initial transient stage. If they are kept connected, the capacitors discharge slowly until thermal equilibrium is reached. If the diodes have different temperatures, the sum of the average charges of the capacitors is not zero even when reaching the final stationary state; see Eq.~\eqref{eq20}.

How is does the system of Figure \ref{fig1} evolve after the initial stage? The time dependent marginal probability density is the product of the stationary marginal probability and two Gaussian functions patched at equal capacitor charges. The average values of the Gaussians consist of two wave fronts advancing towards positive and towards negative charge differences, respectively, cf Figures \ref{fig6} and \ref{fig7}. {\em These waves leave behind them the final stationary state}. The wave fronts are symmetric for the single temperature case and asymmetric otherwise. The variances of the marginal state give the shape of the forefront of the waves and become smaller as the fronts advance. A peculiarity of the charge wave fronts is that their speed becomes exponentially small, they freeze, and it would take an exponentially long time for them to reach infinity. However, the stationary marginal density is so sharp that the effect of the frozen fronts is inappreciable after a finite time; see Fig.~\ref{fig7}.

\acknowledgments
This work has been supported by the FEDER/Ministerio de Ciencia, Innovaci\'on y Universidades -- Agencia Estatal de Investigaci\'on (MCIN/ AEI/10.13039/501100011033) grants PID2020-112796RB-C22 and PID2024-155528RB-C22. This work was financially supported, in part, by a grant from the WoodNext Foundation (AWD-104384), which is administered by the Greater Houston Community Foundation.

\appendix 
\section{Initial charging stage}\label{ap:a}
The initial stage involves the fast time $s=\tau/\epsilon$ and telescoped charge difference $\chi=\xi/\epsilon$ \cite{thi23}. For diodes conducting in opposite directions, the arguments of the conductance functions in Eq.~\eqref{eq3e} are $\pm\eta-\epsilon\chi$, 
\begin{eqnarray}
\mu_1+\mu_2= 1+ O(\epsilon),\,\, \mu_1-\mu_2=-\tanh\frac{\eta}{2w}+O(\epsilon).\quad \label{eqa1}
\end{eqnarray}
\begin{widetext}
Omitting terms of order $\epsilon$, the FPE \eqref{eq3a} becomes 
\begin{subequations}\label{eqa2}
\begin{eqnarray}
\frac{\partial\rho^{(0)}}{\partial s}\!&=&\!\frac{\partial}{\partial\eta}\!\left[(\mu_1+\mu_2)\eta+(\theta_1\mu_1+\theta_2\mu_2) \frac{\partial}{\partial \eta} +(\theta_1\mu_1-\theta_2 \mu_2) \frac{\partial}{\partial\chi} \right]\!\rho^{(0)}\nonumber\\
\!&+&\!\frac{\partial}{\partial\chi}\!\left[(\mu_1-\mu_2)\eta + (\theta_1\mu_1-\theta_2\mu_2)\frac{\partial}{\partial \eta} + (\theta_1\mu_1+\theta_2\mu_2)\frac{\partial}{\partial\chi}\right]\!\rho^{(0)}.\label{eqa2a}
\end{eqnarray}
It is convenient to extract a prefactor from $\rho^{(0)}$,
\begin{eqnarray}
&&\rho^{(0)}= S(\eta)\, R^{(0)}(\eta,\chi,s), \label{eqa2b}\\
&&S(\eta)=\frac{\exp\!\left[-\!\int_0^\eta\!\frac{\mu_1-\mu_2}{\theta_1\mu_1\!-\!\theta_2\mu_2}\eta\,d\eta\right]}{\int_{-\infty}^\infty\exp\!\left[-\!\int_0^\eta\!\frac{\mu_1-\mu_2}{\theta_1\mu_1\!-\!\theta_2\mu_2}\eta\,d\eta\right]d\eta}\label{eqa2c}\\
&&\quad\quad = \frac{e^{-\frac{\eta^2}{2}}\exp\!\left[-(1-\theta_2) \int_0^\eta\!\frac{2y\cosh\frac{y}{2w}}{\theta_2e^\frac{y}{2w}-\theta_1e^{-\frac{y}{2w} }} dy\right]}{\int_0^\infty e^{-\frac{\eta^2}{2}}\!\left[e^{-(1-\theta_2)\int_0^\eta\frac{2y\cosh\frac{y}{2w}dy}{\theta_2 e^\frac{y}{2w} -\theta_1e^{-\frac{y}{2w}} }}+ e^{(1-\theta_2)\int_0^\eta\frac{2y\cosh\frac{y}{2w}dy}{\theta_1 e^\frac{y}{2w} -\theta_2e^{-\frac{y}{2w}} }} \right] d\eta}.\nonumber 
\end{eqnarray}
We have $S(\eta)\propto e^{-\frac{\eta^2}{2\theta_j}}$ as $\eta\to (-1)^j\infty$, $j=1,2$, and $\rho^{(0)}=e^{-\eta^2/2}R^{(0)}(\chi,\eta,s)/\sqrt{2\pi}$ if  $\theta_1=\theta_2 =1$. Substituting Eq.~\eqref{eqa2b} into \eqref{eqa2a} and integrating the result with respect to $\eta$, we get 
\begin{eqnarray}
&&\frac{\partial R^{(0)}}{\partial s} =\frac{\partial}{\partial\chi}\int_{-\infty}^\infty S(\eta)\!\left[(\theta_1\mu_1+\theta_2\mu_2)\frac{\partial}{\partial\chi}+(\theta_1\mu_1-\theta_2\mu_2) \frac{\partial}{\partial \eta}\right]R^{(0)}\,d\eta.\label{eqa2d}
\end{eqnarray}
\end{subequations}
\end{widetext}
Assuming that $R^{(0)}$ is a function of $s$ and of the new variable $\sigma=\chi-\int (\theta_1\mu_1-\theta_2\mu_2)d\eta$, Eq.~\eqref{eqa2d} becomes the heat equation:
\begin{subequations}
\begin{eqnarray}
&&\frac{\partial R^{(0)}}{\partial s}=a\,\frac{\partial^2 R^{(0)}}{\partial\sigma^2}, \label{eqa3a}\\
&&\sigma=\chi+2w\ln\!\left(2\cosh\frac{\eta}{2w}\right)\!+\frac{\theta_2-\theta_1}{2}\eta,\label{eqa3b}\\
&& a= \int_{-\infty}^\infty \frac{4\theta_1\theta_2S(\eta)}{2+\theta_1e^\frac{\eta}{w}+\theta_2e^{-\frac{\eta}{w}}}\, d\eta.\label{eqa3c}
\end{eqnarray}\end{subequations}
For a delta-function initial condition corresponding to the initial zero charge in the circuit, the Gaussian kernel solves this equation and produces the normalized solution
\begin{widetext}
\begin{eqnarray}
\rho^{(0)}(\chi,\eta,s)= \frac{S(\eta)}{2\sqrt{\pi a s}} \exp\!\left[-\frac{\left(\chi\!+\!2w\ln\cosh\!\left(2\frac{\eta}{2w}\right)+\frac{\theta_2-\theta_1}{2}\eta\right)^2}{4as}\right]\!.  \label{eqa4}
\end{eqnarray}
This probability density yields the averages:
\begin{subequations}\label{eqa5}
\begin{eqnarray}
&&\langle\sigma\rangle=0,\quad\langle\sigma^2\rangle=2a s,\quad \langle\eta\rangle=\int_{-\infty}^\infty S(\eta)\eta d\eta\underbrace{=}_{\theta_i=1}0,\label{eqa5a}\\
&&\langle\chi\rangle=-\frac{\theta_2-\theta_1}{2}\langle\eta\rangle-2w \int_{-\infty}^\infty S(\eta)\,\ln\!\left(2\cosh\frac{\eta}{2w}\right) d\eta\underbrace{\sim}_{w\to 0,\theta_i=1} -\sqrt{\frac{2}{\pi}}.\quad\quad\label{eqa5b}
\end{eqnarray}\end{subequations}
\end{widetext}

We can calculate the average energy rate and the entropy production from Eqs.~\eqref{eqa2a} and \eqref{eqa4} using integration by parts. The results are:
\begin{eqnarray}
&&\frac{d}{ds}\langle\mathcal{H}\rangle= O(\epsilon),\nonumber\\
&& \frac{d\Sigma}{ds}=-\frac{d}{ds}\int \rho\,\ln\rho\, d\eta\,d\chi=\frac{1}{as} + O(\epsilon).      \label{eqa6} 
\end{eqnarray}
The production of entropy declines as time elapses. Thus, the entropy increases to a large value after $t=0$ and then it increases  logarithmically as $s\to\infty$ at the end of the initial stage. 

If we go back to the slow scales $\tau$ and $\xi$ of Eq.~\eqref{eq2}, Eq.~\eqref{eqa4} becomes
\begin{eqnarray}
\rho^{(0)}\sim S(\eta)\,\delta(\xi), \label{eqa7}
\end{eqnarray}
as $\epsilon\to 0$. This value will become the initial condition for the reduced equation that  describes the evolution of the probability density in the slow time scale.

 \section{Calculation of $F_1$ in the reduced FPE}\label{ap:b}
 To calculate $\tilde{\rho}^{(1)}$, we use an Ansatz similar to Eqs.~\eqref{eq9} for $\rho^{(0)}$:
\begin{widetext}
\begin{subequations}\label{eqb1}
\begin{eqnarray}
\tilde{\rho}^{(1)}= e^{-\frac{\theta_1-\theta_2}{2(1+\epsilon)}\mathcal {J}(\eta,\xi)}\,\frac{R(\eta,\xi,\tau)}{\overline{E}},\quad\rho^{(1)}=\frac{e^{-\frac{\xi^2}{2\epsilon}}}{2\pi\sqrt{\epsilon(1+\epsilon)}}\hat{E}(\eta,\xi)\, R(\eta,\xi,\tau). \label{eqb1a}
\end{eqnarray}
Then  Eq.~\eqref{eq12b} becomes
\begin{eqnarray}
&&(1+\epsilon)\frac{\partial}{\partial\eta}e^{-\frac{\xi^2}{2\epsilon}}\!\left[(1+\epsilon)(\theta_1\mu_1+\theta_2\mu_2)\hat{E}\frac{\partial R}{\partial\eta}+(\theta_1\mu_1-\theta_2 \mu_2)\frac{\partial(Q\hat{E})}{\partial\xi}\right]\! \nonumber \\
&&\quad=(\theta_1\!-\!\theta_2)\!\left[\hat{E}\frac{\partial}{\partial\xi}(P\, \overline{M})-\frac{\partial}{\partial\xi}(P\,M)\right]\!=(\theta_1\!-\!\theta_2)\!\left[\frac{\partial}{\partial\xi}(P\overline{M}\hat{E}-PM)-P\,\overline{M}\frac{\partial\hat{E}}{\partial\xi}\right]\!. \quad \label{eqb1b}
\end{eqnarray}
Integrating with respect to $\eta$, we get
\begin{eqnarray}
&&(1\!+\!\epsilon)e^{-\frac{\xi^2}{2\epsilon}}\!\!\left[\!(1\!+\!\epsilon)(\theta_1\mu_1\!+\!\theta_2\mu_2)\hat{E}\frac{\partial R}{\partial\eta}\!+\!(\theta_1\mu_1\!-\!\theta_2 \mu_2) \frac{\partial(Q\hat{E})}{\partial\xi}-J(\xi)\right]\! \nonumber\\
&&\quad=\!(\theta_1\!-\!\theta_2)\!\left[\tilde{E}\frac{\partial}{\partial\xi}(P\, \overline{M})-\frac{\partial}{\partial\xi}(P\,\tilde{M})\right]\!=(\theta_1\!-\!\theta_2)\!\left(\Phi\frac{\partial P}{\partial\xi}+\Psi\,P\right)\!, 
\label{eqb1c}\\
&& \tilde{E}=\int \hat{E}d\eta,\quad \tilde{M}=\int M d\eta,\label{eqb1d}\\
&&\Phi=\tilde{E}\overline{M}-\tilde{M},\quad\Psi=\tilde{E}\frac{\partial\overline{M}}{\partial\xi}-\frac{\partial\tilde{M}}{\partial\xi}.  \label{eqb1e}
\end{eqnarray}
\end{subequations}
Here $J(\xi)$ is a probability current and $\tilde{E}$ and $\tilde{M}$ are known up to functions of $\xi$. These functions will be determined so that the integrals appearing in all subsequent formulas converge. 
Eq.~\eqref{eqb1c} yields
\begin{subequations}\label{eqb2}
\begin{eqnarray}
&&\frac{\partial R}{\partial\eta}=e^{\frac{\xi^2}{2\epsilon}}\frac{\frac{(\theta_1-\theta_2)\Phi}{(1+\epsilon)\hat{E}}-(\theta_1\mu_1-\theta_2\mu_2)}{(1+\epsilon)(\theta_1\mu_1+\theta_2\mu_2)}\frac{\partial P}{\partial\xi}+\frac{\frac{(\theta_1-\theta_2)\Psi}{(1+\epsilon)\hat{E}}-(\theta_1\mu_1-\theta_2\mu_2)\!\left(\frac{\partial}{\partial\xi}\ln\hat{E}+\frac{\xi}{\epsilon}\right)\!}{(1+\epsilon)(\theta_1\mu_1+\theta_2\mu_2)}Q\nonumber\\
&&\quad\quad +\frac{J(\xi)}{(1+\epsilon)(\theta_1\mu_1+\theta_2\mu_2)\hat{E}},   \label{eqb2a}\\
&&\rho^{(1)}(\eta,\xi;P)= \frac{e^{-\frac{\xi^2}{2\epsilon}}\hat{E}(\eta,\xi)}{2\pi\sqrt{\epsilon(1+\epsilon)}}\!\left[\int_0^\eta \frac{\partial R}{\partial\eta'}d\eta'-\int_{-\infty}^\infty  \hat{E}(\eta',\xi)\!\left(\int_0^{\eta'} \frac{\partial R}{\partial\eta''}d\eta'' \right)\!d\eta'\right]\!.   \label{eqb2b}
\end{eqnarray}
Redefining the arbitrary function $J(\xi)$ in terms of new functions $\Phi_s(\xi)$ and $\Psi_s(\xi)$, we may rewrite Eq.~\eqref{eqb2a} as
\begin{eqnarray}
\frac{\partial R}{\partial\eta}=e^{\frac{\xi^2}{2\epsilon}}\frac{\frac{(\theta_1\!-\!\theta_2)(\Phi\!-\!\Phi_s)}{(1+\epsilon)\hat{E}}-(\theta_1\mu_1\!-\!\theta_2\mu_2)}{(1+\epsilon)(\theta_1\mu_1+\theta_2\mu_2)}\frac{\partial P}{\partial\xi}+\frac{\frac{(\theta_1\!-\!\theta_2)(\Psi\!-\!\Psi_s)}{(1+\epsilon)\hat{E}}-(\theta_1\mu_1\!-\!\theta_2\mu_2)\!\left(\frac{\partial}{\partial\xi}\ln\hat{E}+\frac{\xi}{\epsilon}\right)\!}{(1+\epsilon)(\theta_1\mu_1+\theta_2\mu_2)}Q.\quad\quad   \label{eqb2c}
\end{eqnarray}
\end{subequations}
The integrals in Eqs.~\eqref{eqb1d} and \eqref{eqb1e} are defined up to arbitrary functions of $\xi$. Eqs.~\eqref{eq16} imply that $\Phi$ and $\Psi$ vanish as $\eta\to\pm \infty$. The additional functions $\Phi_s$ and $\Psi_s$ will be selected so that terms of order $1/\hat{E}$ in Eqs.~\eqref{eq15c}-\eqref{eq15e} cancel out. This will cause the  integrals appearing in the coefficients $\mathcal{B}$ and $\mathcal{D}$ to be convergent. Then Eq.~\eqref{eq13b} produces
\begin{subequations}\label{eqb3}
\begin{eqnarray}
&&F_1\!=\frac{\partial}{\partial\xi}\!\left[A(\xi) \frac{\partial P}{\partial\xi} +B(\xi)P\right]\!,\quad \label{eqb3a}\\
&&A(\xi)\!=\!4\theta_1\theta_2\overline{\frac{\mu_1\mu_2\hat{E}}{\theta_1\mu_1\!+\!\theta_2\mu_2}}\! +\!\frac{\theta_1\!-\!\theta_2}{1+\epsilon} \!\left\{ \overline{\frac{\theta_1\mu_1\!-\!\theta_2\mu_2}{\theta_1\mu_1\!+\!\theta_2\mu_2}\varphi+\frac{M}{\theta_1\mu_1\!+\!\theta_2\mu_2}\!\int_0^\eta\!\!    \frac{\frac{(\theta_1\!-\!\theta_2)\varphi}{(1+\epsilon)\hat{E}}\!-\!\theta_1\mu_1\!+\!\theta_2\mu_2 }{\theta_1\mu_1+\theta_2\mu_2 } }\right.\nonumber\\\!
&&\quad\quad \left.-\overline{M}\,\overline{\hat{E} \int_0^\eta \frac{\frac{\theta_1-\theta_2)\varphi}{(1+\epsilon)\hat{E}}-\theta_1\mu_1+\theta_2\mu_2 }{\theta_1\mu_1+\theta_2\mu_2 } }\right\}, \label{eqb3b}\\
&&B(\xi)\!=4\theta_1\theta_2\overline{\frac{\mu_1\mu_2\hat{E}}{\theta_1\mu_1+\theta_2\mu_2}\! \left(\frac{\partial\ln\hat{E}}{\partial\xi} +\frac{\xi}{\epsilon}\right)\! } +\!\frac{\theta_1\!-\!\theta_2}{1+\epsilon}\left[\overline{\frac{(\theta_1\mu_1\!-\!\theta_2\mu_2)\psi}{\theta_1\mu_1\!+\!\theta_2\mu_2} }+\overline{ \frac{M\, N}{\theta_1\mu_1\!+\!\theta_2\mu_2}}\right]\!, 
\label{eqb3c}\\
&&N\!=\!\int_0^\eta\!  \frac{\frac{\theta_1\!-\!\theta_2}{1+\epsilon}\psi\!-\!(\theta_1\mu_1\!-\!\theta_2\mu_2)\!\!\left(\frac{\partial\hat{E}}{\partial\xi} \!+\!\frac{\hat{E}\xi}{\epsilon}\right)\!}{(\theta_1\mu_1\!+\!\theta_2\mu_2) \hat{E}}d\eta\!-\!\overline{\hat{E} \!\int_0^\eta\!  \frac{\frac{\theta_1\!-\!\theta_2}{1+\epsilon}\psi\!-\!(\theta_1\mu_1\!-\!\theta_2\mu_2)\!\!\left(\frac{\partial\hat{E}}{\partial\xi}\!+\!\frac{\hat{E}\xi}{\epsilon}\right)\!}{(\theta_1\mu_1\!+\!\theta_2\mu_2) \hat{E}}d\eta},\quad\,\label{eqb3d}\\
&&\varphi(\eta,\xi)=\Phi(\eta,\xi)-\Phi_s(\xi),\quad \psi(\eta,\xi)=\Psi(\eta,\xi)-\Psi_s(\xi).  \quad\label{eqb3e}
\end{eqnarray}\end{subequations}

Lastly, substituting Eqs.~\eqref{eq14} and \eqref{eqb3} into \eqref{eq11b}, we get the reduced FPE:
\begin{eqnarray}
\frac{\partial P}{\partial\tau}= \frac{\partial}{\partial\xi}\!\left[\epsilon A(\xi)\frac{\partial P}{\partial\xi}+[(\theta_1-\theta_2)\overline{M}(\xi)+\epsilon B(\xi)]\,P\right]\!,\quad P(\xi,s)=e^{-\frac{\xi^2}{2\epsilon}} Q(\xi,s), \label{eqb4}\end{eqnarray}
Eq.~\eqref{eqb4} is the same as Eq.~\eqref{eq15}.
\end{widetext}

\section{Limit as $w\to 0+$}\label{ap:c}
For the particular case of steplike diode mobility of Eq.~\eqref{eq3e}, we can simplify our formulas for the reduced FPE \eqref{eq15} in the limit of small reference voltage $w$. As $w\to 0+$, we have 
\begin{eqnarray}
&&\mu_1\!\sim\! e^{-\frac{\eta+\xi}{w}}\Theta(\xi)\!+\!\Theta(-\xi), \quad\mu_2\!\sim\! \Theta(\xi)\!+\!\Theta(-\xi) e^{\frac{\xi-\eta}{w}}\!, \quad\label{eqc1} \\
&&\Theta(x)=\!\left\{\begin{array}{cc} 1,& x>0,\\ \frac{1}{2},& x=0,\\ 0,& x<0.\end{array} \right. \nonumber
\end{eqnarray}

\subsection{Gaussian approximation}
From Eq.~\eqref{eq3e} and after some algebra, we get
\begin{widetext}
\begin{subequations}\label{eqc2}
\begin{eqnarray}
&&E(\eta,\xi)=\exp\!\left\{-\frac{1}{1\!+\!\epsilon}\!\int_0^\eta\!\!\left(\eta-\frac{\theta_1\!-\!\theta_2}{2}\xi\right)\!\left(1+\frac{\theta_1\!-\!\theta_2}{2}\frac{e^\frac{\eta}{w} (e^\frac{\xi}{w}-e^{-\frac{\xi}{w}}) }{2+e^\frac{\eta}{w}(\theta_2e^\frac{\xi}{w}+\theta_1e^{-\frac{\xi}{w}})}\right)d\eta\right\}\!\quad \nonumber\\
&&\quad\quad\quad\,=e^{\frac{(\theta_1-\theta_2)^2\xi^2}{8(1+\epsilon)}-\frac{(\eta-\frac{\theta_1-\theta_2}{2}\xi)^2}{2(1+\epsilon)}}\!\exp\!\left\{-\frac{\theta_1-\theta_2}{2(1\!+\!\epsilon)}\!\int_0^\eta\!\!\left(\eta-\frac{\theta_1\!-\!\theta_2}{2}\xi\right)\! \frac{e^\frac{\eta}{w} (e^\frac{\xi}{w}-e^{-\frac{\xi}{w}}) }{2+e^\frac{\eta}{w}(\theta_2e^\frac{\xi}{w}+\theta_1e^{-\frac{\xi}{w}})} d\eta\right\}\!\!,\quad \quad\label{eqc2a}\\
&&\hat{E}(\eta,\xi)=\frac{E(\eta,\xi)}{\overline{E(\eta,\xi)}}.\label{eqc2b}
\end{eqnarray}
The integral in Eq.~\eqref{eqc2a} can be explicitly calculated using $\theta_1+\theta_2=2$ with the result
\begin{eqnarray}
&&\!\int_0^\eta\! \frac{(\eta\!-\! \frac{\theta_1\!-\!\theta_2}{2}\xi)e^\frac{\eta}{w} (e^\frac{\xi}{w}\!-e^{-\frac{\xi}{w}}) }{2+e^\frac{\eta}{w}(\theta_2e^\frac{\xi}{w}+\theta_1e^{-\frac{\xi}{w}})} d\eta= \! \frac{w^2(e^\frac{2\xi}{w}\!-\!1)}{\theta_1+\theta_2e^\frac{2\xi}{w}}\left\{- \frac{\theta_1-\theta_2}{2w}\xi\ln\!\left[\frac{2+\theta_1e^\frac{\eta-\xi}{w}+\theta_2 e^\frac{\eta+\xi}{w} }{(\theta_1+\theta_2e^\frac{\xi}{w})(1-e^{-\frac{\xi}{w}})}\right]\!\right. \quad\nonumber\\
&&\quad+\frac{\eta}{w}\ln\!\left[1\!+\!\frac{e^\frac{\eta-\xi}{w}}{2}(\theta_1\!+\!\theta_2e^\frac{2\xi}{w})\right]\!\!+\!\! \mbox{ Li}_2\!\left[-\frac{\theta_1 e^\frac{\eta-\xi}{w}\!+\!\theta_2e^\frac{\eta+\xi}{w}}{2} \right]\! 
\left.- \mbox{Li}_2\!\left[-\frac{\theta_1e^{-\frac{\xi}{w}}+\theta_2e^\frac{\xi}{w}}{2}\right]\! \right\}\!, \label{eqc2c}
\end{eqnarray}
where Li$_2(z)=\sum_{k=1}^\infty z^k/k^2$ is the polylogarithm function of index 2 (dilogarithm) \cite{polylog}. 
\end{subequations}

Eq.~\eqref{eqc2c} can be approximated as $w\to 0+$ using the formula Li$_2(x)\sim -[\ln (-x)]^2/2$, $|x|\to\infty$. However, the same result is obtained by direct approximation of the integral in Eq.~\eqref{eqc2a}, which is better for our purposes. As $w\to 0+$, we have $e^\frac{\eta+\xi}{w}\!-e^{\frac{\eta-\xi}{w}}\sim$ sign$(\xi)e^\frac{\eta+|\xi|}{w}$, $\theta_2e^\frac{\eta+\xi}{w}+\theta_1e^{\frac{\eta-\xi}{w}}\sim \theta_j e^\frac{\eta+|\xi|}{w}$ for $\eta+|\xi|>0$, and $\int_0^\eta=\int_0^{-|\xi|}+\int_{-|\xi|}^\eta$  for $\eta+|\xi|<0$, which produce
\begin{eqnarray*}
\!\int_0^\eta\! \frac{(\eta\!-\! \frac{\theta_1\!-\!\theta_2}{2}\xi)e^\frac{\eta}{w} (e^\frac{\xi}{w}\!-e^{-\frac{\xi}{w}}) }{2+e^\frac{\eta}{w}(\theta_2e^\frac{\xi}{w}+\theta_1e^{-\frac{\xi}{w}})} d\eta\sim \frac{\mbox{sign}(\xi)}{2\theta_j}\!\left[(2\!-\!\theta_j)\xi^2\Theta(-\eta\!-\!|\xi|)+\eta(\eta-(\theta_1\!-\!\theta_2)\xi)\Theta(\eta\!+|\xi|)\right]  \nonumber\\
= \! \frac{\mbox{sign}(\xi)}{2\theta_j}\!\left\{ \Theta(-\eta\!-\!|\xi|)\xi^2\!\left[\! (2\!-\!\theta_j)^2-\frac{(\theta_1\!-\!\theta_2)^2}{4}\right]\! \! +\!  \Theta(\eta\!+\!|\xi|) \!\left[\!\left(\! \eta\! -\frac{\theta_1\!-\!\theta_2}{2}\xi\right)^2\! -\frac{(\theta_1\!-\!\theta_2)^2\xi^2}{4}\right]\!\right\}\!,
\end{eqnarray*}
where $j=2$ for $\xi>0$ and $j=1$ for $\xi<0$. We have omitted terms of order $w$. Exactly the same result is obtained from Eq.~\eqref{eqc2c}. Using the identity $2\theta_j+(\theta_1-\theta_2)$ sign$(\xi)=2$, $E(\eta,\xi)$ becomes
\begin{subequations}\label{eqc3}
\begin{eqnarray}
E(\eta,\xi)\!\sim \!\sum_{j=1}^2\!\Theta_j e^{\frac{(\theta_1\!-\!\theta_2)^2\!\xi^2}{8(1\!+\!\epsilon)\theta_j}}\!\!\left[\Theta(-\eta\!-\!|\xi|)e^{-\frac{(1\!-\!\theta_j)(2\!-\!\theta_j)^2\xi^2}{2\theta_j(1\!+\!\epsilon)}-\frac{(\eta-\frac{\theta_1\!-\!\theta_2}{2}\xi)^2}{2(1\!+\!\epsilon)}}\!\!+\!\Theta(\eta\!+\!|\xi|) e^{-\frac{(\eta-\frac{\theta_1\!-\!\theta_2}{2}\xi)^2}{2(1\!+\!\epsilon)\theta_j}}\right]\!\!, \quad\quad\, \label{eqc3a} 
\end{eqnarray}
where $\Theta_j=\Theta((-1)^j\xi)$ and $E(0,\xi)=1$. Thus the function $E(\eta,\xi)$ is given by two different Gaussian functions for $\eta<-|\xi|$ and for $\eta>-|\xi|$ if $\theta_1\neq\theta_2$. However, $E(\eta,\xi)$ is continuous at $\eta=-|\xi|$ with
\begin{eqnarray}
E(-|\xi|,\xi)\sim \sum_{j=1}^2\Theta_j e^{\frac{(\frac{\theta_1-\theta_2}{2})^2\xi^2- (2-\theta_j)^2\xi^2}{2(1+\epsilon)\theta_j}}=\sum_{j=1}^2\Theta_j e^{-\frac{(3-2\theta_j) \xi^2}{2(1+\epsilon)\theta_j}}. \label{eqc3b} 
\end{eqnarray}
We have omitted narrow boundary layers at $\xi=0$ and at $\eta=-|\xi|$, whose contributions to the integrals in Eqs.~\eqref{eq15} are of order $w$, negligible as $w\to 0+$. Eq.~\eqref{eqc3a} is an outer approximation so that $|\xi|$ and $|\eta+|\xi||$ are much larger than $w$. Consequently, terms of order $w$ will be neglected in our approximations. We obtain
\begin{eqnarray}
\overline{E}\!\sim\! \sqrt{\!\frac{(1\!+\!\epsilon)\pi}{2}}\!\sum_{j=1}^2\!\Theta_je^{\frac{(\theta_1\!-\!\theta_2)^2\!\xi^2}{8(1\!+\!\epsilon)\theta_j}}\!\!\!\left[e^{\!-\frac{(1\!-\!\theta_j)(2\!-\!\theta_j)^2\!\xi^2}{2\theta_j(1\!+\!\epsilon)}}\!\mbox{erfc}\!\!\left(\!\frac{(2\!-\!\theta_j)|\xi|}{\sqrt{2(1\!+\!\epsilon)}}\!\right)\!\!+\!\!\sqrt{\theta_j}\mbox{erfc}\!\!\left(\!\!-\frac{(2\!-\!\theta_j)|\xi|}{\sqrt{2(1\!+\!\epsilon)\theta_j}}\!\right)\!\!\right]\!\!. \,\,\,\,   \label{eqc3c} 
\end{eqnarray}
Eqs.~\eqref{eqc3a} - \eqref{eqc3c} are independent of $w$. We shall write $\hat{E}(\eta)=\hat{E}_j(\eta,\xi)$ and omit $ \sum_{j=1}^2\Theta_j$ for the sake of simplicity whenever the meaning is clear. 
Similarly, we can approximate $\partial\hat{E}/\partial\xi$ and calculate $d\overline{E}/d\xi$ from the exact Eqs.~\eqref{eqc2}. We obtain
\begin{eqnarray}
&&\frac{\partial\hat{E}}{\partial\xi}\sim-\frac{\theta_1\!-\!\theta_2}{2}\frac{\partial\hat{E}}{\partial\eta}\!+\!\left[(1\!-\!\theta_j)^2 \!-\!(1\!-\!\theta_j)(2\!-\!\theta_j)^2\Theta(-\eta-|\xi|)\right]\!\frac{\xi\,\hat{E}}{(1\!+\!\epsilon)\theta_j}-\frac{\hat{E}}{\overline{E}}\frac{\partial\overline{E}}{\partial\xi}, \label{eqc3d}\\
&&\frac{\partial\overline{E}}{\partial\xi}\sim\!\frac{\theta_1\!-\!\theta_2}{2} e^{\frac{(\theta_1\!-\!\theta_2)^2}{8(1\!+\!\epsilon)\theta_j}\xi^2}|\xi| \sqrt{\frac{\pi}{2(1\!+\!\epsilon)}} \!\left\{\! \!\left(\! 1\! -\! \theta_j\!-\!\frac{3\! -\! 2\theta_j}{\theta_j}\!\right)\! e^{-\frac{(1\!-\!\theta_j)(2-\theta_j)^2\xi^2}{2(1\!+\!\epsilon)\theta_j}} \mbox{erfc}\!\left(\frac{(2\!-\!\theta_j)|\xi|}{\sqrt{2(1\!+\!\epsilon)}}\!\right)\!  \right. \nonumber\\
&&\quad\,\left. + \frac{1\!-\!\theta_j}{\sqrt{\theta_j}}  \mbox{erfc}\!\left(-\frac{(2\!-\!\theta_j)|\xi|}{\sqrt{2(1\!+\!\epsilon)\theta_j}}\right)\!\right\}\nonumber\\
&&\quad\sim \frac{(1\!-\!\theta_j)\overline{E}}{(1\!+\!\epsilon)\theta_j}\xi\!\left[1\!-\!\theta_j-\sqrt{\frac{\pi(1\!+\!\epsilon)}{2}} (2\!-\!\theta_j)^2\hat{E}(-|\xi|,\xi)\,e^\frac{(2\!-\!\theta_j)^2\xi^2}{2(1+\epsilon)}\! \mbox{erfc}\!\left(\frac{(2\!-\!\theta_j)|\xi|}{\sqrt{2(1\!+\!\epsilon)}}\!\right)\!\right]\!,\label{eqc3e}
\end{eqnarray}
\end{subequations}
which also follows from direct differentiation of Eq.~\eqref{eqc3c}. 

\subsection{Calculation of $M$, and choice of $\Phi$, $\Psi$}
$M(\eta,\xi)$, $\int_\eta^\infty M\,d\eta$ and $\overline{M}$ in Eqs.~\eqref{eq14} become
\begin{subequations}\label{eqc4}
\begin{eqnarray}
&&M=\frac{(\theta_1-\theta_2)\xi-2\eta}{\frac{\theta_1}{\mu_2}+\frac{\theta_2}{\mu_1}}\hat{E}=\frac{(\theta_1-\theta_2)\xi-2\eta}{2+\theta_1 e^\frac{\eta-\xi}{w}+\theta_2 e^\frac{\eta+\xi}{w}}\hat{E}\nonumber\\ 
&&\quad \sim -2\!\left(\eta-\frac{\theta_1-\theta_2}{2}\xi\right)\!\left[ \frac{1}{\theta_j} e^{-\frac{\eta+|\xi|}{w}}\Theta(\eta+|\xi|)+\frac{\Theta(-\eta-|\xi|)}{2}\right]\hat{E}\Longrightarrow\nonumber\\ 
&&M\sim 2(1+\epsilon)\!\left[e^{-\frac{\eta+|\xi|}{w}}\Theta(\eta+|\xi|)+\frac{1}{2}\Theta(-\eta-|\xi|)\right]\!\frac{\partial\hat{E}}{\partial\eta}, \label{eqc4a}\\
&&\overline{M} \sim (1\!+\!\epsilon)\!\left[\hat{E}(-|\xi|)+2\!\!\int_{-|\xi|}^\infty\! e^{-\frac{\eta+|\xi|}{w}}\frac{\partial\hat{E}}{\partial\eta}d\eta\right]\!
\sim(1\!+\!\epsilon)\hat{E}(-|\xi|) \sim\frac{1\!+\!\epsilon}{\overline{E}} e^{-\frac{(3-2\theta_j)\xi^2}{2(1+\epsilon)\theta_j}} , \quad\label{eqc4b}\\ 
\!&&\frac{\theta_1\!-\!\theta_2}{2}\xi+|\xi|\!=\!\frac{|\xi|}{2}[2\!+\!(\theta_1\!-\!\theta_2)\mbox{sign}(\xi)]\!=\!|\xi|\!\sum_{j=1}^2(2\!-\!\theta_j)\Theta_j,\, |\xi|\!- \frac{\theta_1\!-\!\theta_2}{2}\xi\!=\!|\xi|\!\sum_{j=1}^2\Theta_j\theta_j.\quad\quad \label{eqc4c}
\end{eqnarray}\end{subequations}
In Eq.~\eqref{eqc4b} we have approximated the integral by using the formulas (Watson's lemma \cite{bender})
\begin{subequations}\label{eqc5}
\begin{eqnarray}
&&\!\int_\eta^\infty\! \! e^{-\frac{y+|\xi|}{w}}f(y)\, dy= e^{-\frac{\eta+|\xi|}{w}}\frac{w}{1-w\frac{\partial}{\partial\eta}}f(\eta),\label{eqc5a}\\
&&\!\int_\eta^\infty\! y e^{-\frac{y}{w}}f(y)\, dy= w^2\frac{\partial}{\partial w} e^{-\frac{\eta}{w}}\frac{w}{1-w\frac{\partial}{\partial\eta}}f(\eta),\label{eqc5b}
\end{eqnarray}
that hold provided the largest value of the integrand occurs at $\eta$ (the lower limit of the integral), $F'(\eta)\neq 0$ and $F(\eta)$ and all its derivatives tend to zero as $\eta\to\infty$. 
\end{subequations}

To approximate $\Phi$ and $\Psi$ in Eq.~\eqref{eqb1d}, we need to find appropriate versions of the indefinite integrals of $M$ and $\hat{E}$, $\tilde{M}$ and $\tilde{E}$ in Appendix \ref{ap:b}, that render finite all integrals in Eqs~\eqref{eq15}. Let us start with $M$. We have
\begin{subequations}\label{eqc6}
\begin{eqnarray}
&&\frac{\int_{-\infty}^\eta\! M\, d\eta}{2(1+\epsilon)}\sim\frac{1}{2} \Theta(-\eta-|\xi|)\hat{E}(\eta) +\frac{\Theta(\eta+|\xi|)}{2(1+\epsilon)}\!\left(\overline{M}-\int_\eta^\infty M\, d\eta\right)\!  \nonumber\\
&&\quad\sim\frac{1}{2} \Theta(-\eta-|\xi|)\hat{E}(\eta) -\Theta(\eta+|\xi|)\int_\eta^\infty e^{-\frac{\eta+|\xi|}{w}}\frac{\partial\hat{E}}{\partial\eta}d\eta +\frac{\overline{M}}{2(1+\epsilon)}\Theta(\eta+|\xi|),\quad\label{eqc6a}\end{eqnarray}
\begin{eqnarray}
&&\frac{\int_\eta^\infty\! M\,d\eta}{2(1+\epsilon)}\sim\Theta(-\eta-|\xi|)\!\left( \!\int_{-|\xi|}^\infty e^{-\frac{\eta+|\xi|}{w}}\frac{\partial\hat{E}}{\partial\eta}d\eta+\frac{1}{2}[\hat{E}(-|\xi|)-\hat{E}(\eta)]\right)   \nonumber\\
&&\quad\quad\quad\quad+\Theta(\eta+|\xi|)\int_\eta^\infty e^{-\frac{\eta+|\xi|}{w}}\frac{\partial\hat{E}}{\partial\eta}d\eta\sim \Theta(\eta+|\xi|)\int_\eta^\infty e^{-\frac{\eta+|\xi|}{w}}\frac{\partial\hat{E}}{\partial\eta}d\eta \nonumber\\
&&\quad\quad\quad\quad -\frac{1}{2} \Theta(-\eta-|\xi|)\hat{E}(\eta) +\frac{\overline{M}}{2(1+\epsilon)}\Theta(-\eta-|\xi|).   \label{eqc6b}
\end{eqnarray}\end{subequations}

Eq.~\eqref{eqb2a} has terms proportional to $1/(\theta_1\mu_1+\theta_2\mu_2)$ that diverge as $\eta\to\pm\infty$ when inserted in Eqs.~\eqref{eq15}. As $w\to 0+$, the last term in Eq.~\eqref{eqb2a} contains the factor 
\begin{eqnarray}
\frac{1}{\theta_1\mu_1\!+\!\theta_2\mu_2}=\frac{1\!+\!e^\frac{2\eta}{w} \!+\! 2 e^\frac{\eta}{w} \cosh\frac{\xi}{w} }{2\!+\!e^\frac{\eta}{w} (\theta_1 e^{-\frac{\xi}{w}}\!+\!\theta_2 e^\frac{\xi}{w})}\sim \frac{\Theta(\xi^2-\eta^2)+e^\frac{\eta-|\xi|}{w}\Theta(\eta-|\xi|)}{\theta_j}+\frac{\Theta(-\eta-|\xi|)}{2},\quad \quad \label{eqc7}
\end{eqnarray}
and it is proportional to the probability current $J(\xi)$. When integrated over $\eta$, divided by $\hat{E}$, integrated again over $\eta$, multiplied by $\hat{E}$ and integrated for $\eta\in\mathbb{R}$, the terms that will become divergent are $e^\frac{\eta-|\xi|}{w}\Theta(\eta-|\xi|)/\theta_j$ as $\eta\to\infty$ and $\Theta(-\eta-|\xi|)/2$ as $\eta\to-\infty$. These terms should compensate the divergences issuing from $\Phi/[(\theta_1\mu_1+\theta_2\mu_2)\hat{E}]$ and $\Psi/[(\theta_1\mu_1+\theta_2\mu_2)\hat{E}]$  in Eq.~\eqref{eqb2a}. In turn, these divergences are proportional to $\overline{M}$ in Eqs.~\eqref{eqc6a} and \eqref{eqc6b}. However, defining $\Phi$ and $\Psi$ as in Eqs.~\eqref{eq16}, 
these functions tend to 0 as $\eta\to\pm\infty$ and do not have a singular part. Thus, using the definitions \eqref{eq16}, we can set $J(\xi)=0$ in Eq.~\eqref{eqb2a}. Integrating by parts, we get
\begin{subequations}\label{eqc8}
\begin{eqnarray}
&&\overline{\Phi}=\frac{1}{2}\!\int_{-\infty}^\infty\!\!\left[ \int_{-\infty}^\eta\!(\overline{M}\hat{E}- M)\,d\eta+\!\int_\eta^\infty\!(M-\overline{M}\hat{E})\,d\eta\right] d\eta\nonumber\\
&&\quad=\frac{1}{2}\!\left[ \eta\!\int_{-\infty}^\eta\!(\overline{M}\hat{E}- M)\,d\eta+\eta\!\int_\eta^\infty\!(M-\overline{M}\hat{E})\,d\eta\right]_{-\infty}^\infty +\!\int_{-\infty}^\infty\!\eta\, (M-\overline{M}\hat{E})\,d\eta\Longrightarrow\nonumber\\
&&\overline{\Phi}= \int_{-\infty}^\infty\!\eta\, (M-\overline{M}\hat{E})\,d\eta.\label{eqc8a}
\end{eqnarray}
Similarly, we get from Eq.~\eqref{eq16b} after integration by parts:
\begin{eqnarray}
\overline{\Psi}=\int_{-\infty}^\infty\!\eta\left(\frac{\partial M}{\partial\xi}-\frac{d\overline{M}}{d\xi}\hat{E}\right) d\eta. \label{eqc8b}
\end{eqnarray}\end{subequations}
Using Eqs.~\eqref{eqc3}, \eqref{eqc4} and \eqref{eqc8} and ignoring $O(w)$ terms, we obtain
\begin{subequations}\label{eqc9}
\begin{eqnarray}
&&\!\overline{\Phi}\!\sim\! (1\!-\!\theta_j)\overline{M}^2-(2-\theta_j)|\xi|\overline{M}-\!(1\!+\!\epsilon)\mathcal{M},\nonumber\\ 
&&\!\overline{\Psi}\!\sim\! \frac{1\!-\!\theta_j}{2}\frac{d\overline{M}^2}{d\xi} \!-\!(2\!-\!\theta_j)|\xi|\frac{d\overline{M}}{d\xi} -(1\!+\!\epsilon)\frac{d\mathcal{M}}{d\xi}-\overline{M}\mbox{sign}\xi.   \label{eqc9a}
\end{eqnarray}
We have used
\begin{eqnarray}
&&\mathcal{M}=\overline{M}\sqrt{\frac{\pi}{2(1\!+\!\epsilon)}} e^\frac{(2\!-\!\theta_j)^2\xi^2}{2(1+\epsilon)}\mbox{erfc}\!\left(\!\frac{(2-\theta_j)|\xi|}{\sqrt{2(1\!+\!\epsilon)}}\right)\!\sim \!\int_{-\infty}^{-|\xi|}\! \hat{E}\, d\eta,\quad \int_{-|\xi|}^\infty\!\hat{E}\, d\eta\sim 1-\mathcal{M},\quad\label{eqc9b}\\
&&\overline{\eta\hat{E}}\!\sim\!  (1\!-\!\theta_j)(|\xi|\!-\!\overline{M}),\quad\overline{\eta M}\!\sim\!-\overline{M} |\xi|\!-\!(1\!+\!\epsilon)\mathcal{M},\label{eqc9c}\\
&&\int_{\infty}^{-|\xi|}\!\eta\hat{E}\,d\eta\sim\! (1\!-\!\theta_j)|\xi|\mathcal{M}-\overline{M},\quad\int_{-|\xi|}^\infty\!\eta\hat{E}\,d\eta\sim\! (1\!-\!\theta_j)|\xi|(1-\mathcal{M})\!+\!\theta_j\overline{M}.\label{eqc9d}
\end{eqnarray}
We also have\begin{eqnarray}
\frac{\partial\overline{E}}{\partial\xi}\!\sim\!\frac{(1\!-\!\theta_j)^2\overline{E}}{\theta_j(1\!+\!\epsilon)}\!\left(\!1\!-\!\frac{(2\!-\!\theta_j)^2}{1\!-\!\theta_j}\mathcal{M\!}\right)\!\xi,   \,\,\frac{dM}{d\xi}\!\sim \!(1\!+\!\epsilon)\frac{d\hat{E}(-|\xi|)}{d\xi}\!\sim\! (1\!+\!\epsilon)\frac{d}{d\xi}\!\left( \frac{e^{-\frac{(3-2\theta_j)\xi^2}{2(1+\epsilon)\theta_j}}}{\overline{E}(\xi)}\right)\!\! .\,\, \label{eqc9e}
\end{eqnarray}
\end{subequations}

\subsection{Calculation of $\mathcal{A}$}
In Eqs.~\eqref{eq15b} and \eqref{eqb2a}, after using Eqs.~\eqref{eqc5}, we have 
\begin{subequations} \label{eqc10}
\begin{eqnarray}
&&\frac{\theta_1\mu_1\!-\!\theta_2\mu_2}{\theta_1\mu_1\!+\!\theta_2\mu_2}=\frac{\theta_1\!-\!\theta_2\!+\!\theta_1e^\frac{\eta-\xi}{w}\!-\!\theta_2e^\frac{\eta+\xi}{w}}{2+e^\frac{\eta}{w}(\theta_1 e^{-\frac{\xi}{w}}+\theta_2e^\frac{\xi}{w})}\sim\! \mbox{ sign}(\xi)[(1\!-\!\theta_j)\Theta(-\eta\!-\!|\xi|) -\Theta(\eta\!+\!|\xi|)], \quad\quad\label{eqc10a}\\
&&\int_0^\eta\frac{\theta_1\mu_1\!-\!\theta_2\mu_2}{\theta_1\mu_1\!+\!\theta_2\mu_2}\sim\int_0^\eta\frac{\theta_1\!-\!\theta_2-\theta_j\mbox{sign}\xi e^\frac{\eta\!+\!|\xi|}{w}}{2\!+\!\theta_je^\frac{\eta\!+\!|\xi|}{w}}\sim \{\Theta(-\eta\!-\!|\xi|)[(1\!-\!\theta_j)\eta\!+\!(2\!-\!\theta_j)|\xi|]\nonumber\\
&& \quad\quad\quad\quad\quad\quad\quad -\, \eta\Theta(\eta\!+\!|\xi|)]\}\mbox{sign}\xi,\label{eqc10b}\\
&&(\theta_1\!-\!\theta_2)\overline{\hat{E}\!\int_0^\eta\frac{\theta_1\mu_1\!-\!\theta_2\mu_2}{\theta_1\mu_1\!+\!\theta_2\mu_2}}\sim 2(1\!-\!\theta_j)\!\left[(2\!-\!\theta_j)^2|\xi|\mathcal{M}-(1\!-\!\theta_j)|\xi|-\overline{M}\right]\!, \label{eqc10c}\\
&&\!\overline{\frac{(\theta_1\!-\!\theta_2)\hat{M}}{\theta_1\mu_1\!+\!\theta_2\mu_2}\!\int_0^\eta\frac{\theta_1\mu_1\!-\!\theta_2\mu_2}{\theta_1\mu_1\!+\!\theta_2\mu_2}}\sim -(1\!-\!\theta_j)\!\left[|\xi|\overline{M} +  (1\!-\!\theta_j)(1\!+\!\epsilon)\mathcal{M} \right]\!,  \label{eqc10d}\\
&&\overline{\frac{\mu_1\mu_2\hat{E}}{\theta_1\mu_1+\theta_2\mu_2}}\sim\int_{-\infty}^\infty\!\left[\frac{\Theta(-\eta-|\xi|)}{2}+\Theta(\eta+|\xi|)\frac{e^{-\frac{\eta+|\xi|}{w}}}{\theta_j}\right] \!\hat{E}(\eta)d\eta\sim \frac{\mathcal{M}}{2}, \label{eqc10e}\\
&&\frac{M}{\theta_1\mu_1\!+\!\theta_2\mu_2}\sim 2(1\!+\!\epsilon)\!\left[\frac{1}{4}\Theta(-\eta-|\xi|)+e^{-\frac{\eta+|\xi|}{w}}\frac{\Theta(\xi^2\!-\!\eta^2)}{\theta_j}+e^{-\frac{2|\xi|}{w}}\frac{\Theta(\eta-|\xi|)}{\theta_j}\right]\!\frac{\partial\hat{E}}{\partial\eta},\label{eqc10f}\\
&&\overline{\frac{M}{\theta_1\mu_1\!+\!\theta_2\mu_2}}\sim \frac{(1\!+\!\epsilon)\hat{E}(-|\xi|)}{2} \sim \frac{\overline{M}}{2}, \quad\quad 
\label{eqc10g}
\end{eqnarray}
 as $|\xi|\gg w$. 
\end{subequations}
Substituting in Eq.~\eqref{eq15b}, we obtain 
\begin{subequations}\label{eqc11}
\begin{eqnarray}
\mathcal{A}\sim\frac{(1-\theta_j)(\theta_j|\xi|-\overline{M})\overline{M}}{1+\epsilon}+\!\left(2\theta_1\theta_2+(1-\theta_j)^2+\frac{(1-\theta_j)(2-\theta_j)^2|\xi|\overline{M}}{1+\epsilon}\right)\mathcal{M}.
\label{eqc11a}
\end{eqnarray}
From Eq.~\eqref{eqc3c}, \eqref{eq4b} and \eqref{eq9b}, we deduce 
\begin{eqnarray}
\overline{M}\sim \!\sqrt{\frac{1+\epsilon}{2\pi\theta_j}}e^{-\frac{(2\!-\!\theta_j)^2\xi^2}{2(1+\epsilon)\theta_j}},\quad\mathcal{M}\sim \!\sqrt{\frac{1+\epsilon}{2\pi\theta_j}} \frac{e^{-\frac{(2\!-\!\theta_j)^2\xi^2}{2(1+\epsilon)\theta_j}}}{(2-\theta_j)|\xi|}\quad \mbox{as } |\xi|\gg 1.  \label{eqc11b}
\end{eqnarray}
Then the products of rapidly decreasing exponentials such as $\overline{M}^2$ and $\overline{M}\mathcal{M}$ can be ignored in Eq.~\eqref{eqc11a}, which becomes 
\begin{eqnarray}
\mathcal{A}\sim  
\sqrt{\frac{(1\!+\!\epsilon)\theta_j}{2\pi}}\!\left(\frac{2}{|\xi|}\!+\!\frac{(1\!-\!\theta_j)|\xi|}{1+\epsilon}\right)\! e^{-\frac{(2-\theta_j)^2\xi^2}{2(1\!+\!\epsilon)\theta_j}} \quad\mbox{as } |\xi|\gg 1. \label{eqc11c}
\end{eqnarray}
\end{subequations}

\subsection{Calculation of $\mathcal{D}$}
Next, we calculate $\mathcal{D}$ in Eq.~\eqref{eq15e}. The function $\Phi$ is
\begin{subequations}\label{eqc12}
\begin{eqnarray}
&&\Phi\sim\Theta(-\eta\!-\!|\xi|)\!\left[\overline{M}\mathcal{M}\frac{\mbox{erfc}\!\left(\!\frac{(1-\theta_j)|\xi|-\eta}{\sqrt{2(1\!+\!\epsilon)}}\right)\! }{\mbox{erfc}\!\left(\!\frac{(2-\theta_j)|\xi|}{\sqrt{2(1\!+\!\epsilon)}}\right)\!}-(1\!+\!\epsilon)\hat{E}\right]\! +\Theta(\eta\!+\!|\xi|)\!\left[2(1\!+\!\epsilon)\!\int_\eta^\infty\!\!e^{-\frac{\eta\!+\!|\xi|}{w}}\frac{\partial\hat{E}}{\partial\eta} d\eta\right.\nonumber\\
&&\quad\quad\left.-\overline{M}(1-\mathcal{M} )\frac{\mbox{erfc}\!\left(\!\frac{\eta-(1-\theta_j)|\xi|}{\sqrt{2(1\!+\!\epsilon)\theta_j}}\right)\!}{\mbox{erfc}\!\left(-\frac{(2-\theta_j)|\xi|}{\sqrt{2(1\!+\!\epsilon)\theta_j}}\right)\!}\right]\!,\quad \quad\label{eqc12a}
\end{eqnarray}
so that 
\begin{eqnarray}
\Phi(-|\xi|)\sim (\mathcal{M}-1)\overline{M},\quad\overline{\Phi}\sim(1\!-\!\theta_j)\overline{M}^2-(2\!-\!\theta_j)|\xi|\overline{M}-(1\!+\!\epsilon)\mathcal{M}.\label{eqc12b}\
\end{eqnarray}
Then
\begin{eqnarray}
&&\overline{\Phi\,\frac{\theta_1\mu_1\!-\!\theta_2\mu_2}{\theta_1\mu_1\!+\!\theta_2\mu_2}}\!\sim\!\left[\overline{M}^2\!-(1\!+\!\epsilon)(1\!-\!\theta_j)\mathcal{M}\right]\!\mbox{sign}\xi\!+\!(2\!-\!\theta_j)[1\!-\!(2\!-\!\theta_j)\mathcal{M}]\overline{M}\xi,\quad\quad\label{eqc12c}\\
&&\!\frac{\frac{\Phi}{(1+\epsilon)\hat{E}}}{\theta_1\mu_1\!+\!\theta_2\mu_2}\sim\frac{\Theta(-\eta\!-\!|\xi|)}{2}\!\left[\frac{\sqrt{\pi}\,\overline{M}}{\sqrt{2(1\!+\!\epsilon)}}e^\frac{[\eta-(1\!-\!\theta_j)\xi]^2}{2(1+\epsilon)}\mbox{erfc}\!\left(\!\frac{(1-\theta_j)|\xi|-\eta}{\sqrt{2(1\!+\!\epsilon)}}\right)\! -1\right]\nonumber\\
&&+\frac{\Theta(\xi^2\!-\!\eta\!^2)}{\theta_j}e^\frac{[\eta-(1\!-\!\theta_j)\xi]^2}{2(1+\epsilon)\theta_j}\!\!\left[\!2\!\!\int_\eta^\infty\!\!e^{-\frac{\eta\!+\!|\xi|}{w}}\frac{\partial e^{-\frac{[\eta-(1\!-\!\theta_j)\xi]^2}{2(1+\epsilon)\theta_j}}}{\partial\eta} d\eta\!-\!\frac{\sqrt{\pi\theta_j}\,\overline{M}}{\sqrt{2(1\!+\!\epsilon)}}\mbox{erfc}\!\left(\!\frac{\eta\!-\!(1\!-\!\theta_j)|\xi|}{\sqrt{2(1\!+\!\epsilon)}}\right)\!\! \right]\! \nonumber\\
&&+\frac{\Theta(\eta\!-\!|\xi|)}{\theta_j}e^{\frac{[\eta-(1\!-\!\theta_j)\xi]^2}{2(1+\epsilon)\theta_j}+\frac{\eta\!-\!2|\xi|}{w}}\!\!\left[\!2\!\!\int_\eta^\infty\!\!e^{-\frac{\eta}{w}}\frac{\partial e^{-\frac{[\eta-(1\!-\!\theta_j)\xi]^2}{2(1+\epsilon)\theta_j}}}{\partial\eta} \!-\!\frac{\sqrt{\pi\theta_j}\,\overline{M}}{\sqrt{2(1\!+\!\epsilon)}}\mbox{erfc}\!\left(\!\frac{\eta\!-\!(1\!-\!\theta_j)|\xi|}{\sqrt{2(1\!+\!\epsilon)}}\right)\!\! \right]\!\!.\quad\quad\quad\label{eqc12d}
\end{eqnarray}
When integrated on $(0,\eta)$, the terms on the second and third line of Eq.~\eqref{eqc12d} produce terms of orders $w$ and $e^{-2|\xi|/w}$ that do not contribute to Eq.~\eqref{eqc13a} below. Then
\begin{eqnarray}
\!\int_0^\eta\!\frac{\frac{\Phi}{(1+\epsilon)\hat{E}}d\eta}{\theta_1\mu_1\!+\!\theta_2\mu_2}\sim\frac{\Theta(-\eta\!-\!|\xi|)}{2}\!\left[\frac{\overline{M}\mathcal{M}}{(1\!+\!\epsilon)\mbox{erfc}\!\left(\!\frac{(2-\theta_j)|\xi|}{\sqrt{2(1\!+\!\epsilon)}}\right)\!}\!\int_{-|\xi|}^\eta\! \frac{\mbox{erfc}\!\left(\!\frac{(1-\theta_j)|\xi|-\eta}{\sqrt{2(1\!+\!\epsilon)}}\right)\! }{\hat{E}}d\eta-\eta\right]\!\!,\quad\label{eqc12e}
\end{eqnarray}
\end{subequations}
and the following terms appear in Eq.~\eqref{eq15e}:
\begin{subequations}\label{eqc13}
\begin{eqnarray}
&&\!\!\overline{\frac{M}{\theta_1\mu_1\!+\!\theta_2\mu_2}\!\int_0^\eta\! \! \frac{\frac{\Phi}{(1+\epsilon)\hat{E}}\, d\eta}{\theta_1\mu_1\!+\!\theta_2\mu_2}}\!\sim\! \frac{\overline{M}(|\xi|\!-\!\overline{M})}{2}+\frac{\mathcal{M}}{2}\!\left(1\!+\!\epsilon+(2\!-\!\theta_j)|\xi|\overline{M}\right)\!, \label{eqc13a}\\
&& \!\!\overline{\hat{E}\!\int_0^\eta\! \! \frac{\frac{\Phi}{(1+\epsilon)\hat{E}}\, d\eta}{\theta_1\mu_1\!+\!\theta_2\mu_2}}\!\sim\!\overline{M}\mathcal{M} \frac{g\!\left(\!\frac{(2-\theta_j)|\xi|}{\sqrt{2(1\!+\!\epsilon)}}\right)\!}{\mbox{erfc}\!\left(\!\frac{(2-\theta_j)|\xi|}{\sqrt{2(1\!+\!\epsilon)}}\right)\!}+(1-\theta_j)\frac{\overline{M}-|\xi|}{2},\quad\quad \label{eqc13b}\\
 &&g(x)=\!\int_x^\infty\! e^{-y^2}\!\int_x^y\! e^{z^2}\mbox{erfc}(z) dz\, dy,\quad g(x)\sim\frac{e^{-x^2}}{4x^2}\,\mbox{ as } x\to\infty.\label{eqc13c}
\end{eqnarray}
\end{subequations}
Note that we can calculate the function $g(x)$ in a different fashion. After finding numerically $g(0)\approx 0.0228097$, we can solve
\begin{subequations}\label{eqc14}
\begin{eqnarray}
&&g'(x)=-e^{x^2}\mbox{erfc}(x)  \int_x^\infty\! e^{-y^2}dy= -\frac{\sqrt{\pi}}{2}e^{x^2}[\mbox{erfc}(x) ]^2,\label{eqc14a}\\
&&g(x)=g(0)-\frac{\sqrt{\pi}}{2}\int_0^xe^{y^2}[\mbox{erfc}(y) ]^2dy,\quad g(x)\sim \frac{\sqrt{\pi}}{|x|} e^{x^2}\,\mbox{ as } x\to-\infty.    \label{eqc14b}
\end{eqnarray}
$g(x)\approx -0.324$ for $x\geq 2$ and it becomes exceedingly large ($10^5$ and rapidly increasing) for $x<-3.5$.
\end{subequations}

The coefficient $\mathcal{D}$ becomes
\begin{subequations}\label{eqc15}
\begin{eqnarray}
&&\mathcal{D}\!\sim\! \frac{1\!-\!\theta_j}{1\!+\!\epsilon}\overline{M}\!\!\left[\![7\!-\!6\theta_j\!+\!\theta_j^2\!-\!2(2\!-\!\theta_j)\mathcal{M}]|\xi| 
\!+\!\overline{M}\!\!\left(\! 4\theta_j\!-\! 1\!-\!\theta_j^2\!-\!2(1\!-\!\theta_j)\frac{\mathcal{M} g\!\left(\!\frac{(2-\theta_j)|\xi|}{\sqrt{2(1\!+\!\epsilon)}}\!\!\right)\! }{\mbox{erfc}\!\left(\!\frac{(2-\theta_j)|\xi|}{\sqrt{2(1\!+\!\epsilon)}}\!\!\right)\!} \!\right)\! \!\right]\!\!.\quad\quad\quad \label{eqc15a}
\end{eqnarray}
As $ |\xi|\gg 1$, Eq.~\eqref{eqc11b} implies that Eq.~\eqref{eqc15a} becomes 
\begin{eqnarray}
 \mathcal{D}\!\sim\! \frac{1-\theta_j}{\sqrt{2\pi\theta_j(1\!+\!\epsilon)}} (7\!-\!6\theta_j\!+\!\theta_j^2)|\xi| e^{-\frac{(2\!-\!\theta_j)^2\xi^2}{2(1+\epsilon)\theta_j}}.
  \label{eqc15b}
\end{eqnarray}
\end{subequations}

\subsection{Calculation of $\mathcal{B}$}
To calculate $\mathcal{B}$ in Eq.~\eqref{eq15c}, we need
\begin{subequations}\label{eqc16}
\begin{eqnarray}
&&\overline{\frac{\mu_1\mu_2}{\theta_1\mu_1+\theta_2\mu_2}\frac{\partial\hat{E}}{\partial\xi}}\sim\int_{-\infty}^\infty\!\left[\frac{\Theta(-\eta-|\xi|)}{2}+\Theta(\eta+|\xi|)\frac{e^{-\frac{\eta+|\xi|}{w}}}{\theta_j}\right] \!\frac{\partial\hat{E}}{\partial\xi}d\eta\sim\frac{1}{2}\!\int _{-\infty}^{-|\xi|} \!\frac{\partial\hat{E}}{\partial\xi}d\eta\nonumber\\
&&\quad+\frac{1}{\theta_j}\!\int _{-|\xi|}^\infty\!e^{-\frac{\eta+|\xi|}{w}} \frac{\partial\hat{E}}{\partial\xi}d\eta\sim -\frac{(\theta_1\!-\!\theta_2)}{2(1\!+\!\epsilon)}\!\left\{\frac{\overline{M}}{2}+\frac{(2\!-\!\theta_j)^2\mathcal{M}(1\!+\!\mathcal{M})|\xi|}{\theta_j}\right\}\!, \label{eqc16a}\\
&&\!\int_0^\eta\!  \frac{\!(\theta_1\mu_1\!-\!\theta_2\mu_2)\frac{\partial\hat{E}}{\partial\xi}}{(\theta_1\mu_1\!+\!\theta_2\mu_2)\hat{E}}d\eta\sim\mbox{sign}\xi\left[\Theta(-\eta\!-\!|\xi|)(1\!-\!\theta_j)\!-\!\Theta(\eta\!+\!|\xi|)\right]\int_0^\eta\!\frac{\partial\ln\hat{E}}{\partial\xi}d\eta\! \sim \!\frac{1-\theta_j}{\theta_j(1\!+\!\epsilon)}\nonumber\\
&&\times\! \left\{\frac{\Theta(\eta\!+\!|\xi|)}{2}[ (1\!-\!\theta_j)^2\xi^2\!+\!2(2\!-\!\theta_j)^2\mathcal{M}\eta\xi\!-\! (\eta\!-\!(1\!-\!\theta_j)|\xi|)^2 ]\!+\! \Theta(-\eta\!-\!|\xi|) (1\!-\!\theta_j)\right. \nonumber\\
&&\left.\!\times \!\left[\frac{\theta_j}{2}(\eta\!-\!(1\!-\!\theta_j)|\xi|)^2\!-\! (2-\theta_j)^2\!\!\left(\!(1\!+\!\mathcal{M})(\eta\!+\!|\xi|)\!- \!\frac{\mathcal{M}|\xi|}{2}\right)\! \xi\!-\! (3\!-\!2\theta_j)\xi^2\right]\!\right\} \!\!, \label{eqc16b}
\end{eqnarray}
\begin{eqnarray}
&&\overline{\frac{M}{\theta_1\mu_1\!+\!\theta_2\mu_2}\!\int_0^\eta\!  \frac{\theta_1\mu_1\!-\!\theta_2\mu_2}{\theta_1\mu_1\!+\!\theta_2\mu_2}\frac{\partial\ln\hat{E}}{\partial\xi}}\sim\frac{\overline{M}(1\!-\!\theta_j)^2\xi^2}{4(1\!+\!\epsilon)^2\theta_j}[3\!-\!2\theta_j\!+\!2(2\!-\!\theta_j)^2\mathcal{M}] \nonumber\\
&&\quad+\frac{\overline{M}(1\!-\!\theta_j)^2}{2(1\!+\!\epsilon)}-\frac{(1\!-\!\theta_j)^2(2\!-\!\theta_j)^2(\mathcal{M}-1)|\xi|}{(1\!+\!\epsilon)\theta_j},\label{eqc16c}\\
&&\overline{\hat{E}\!\int_0^\eta\!  \frac{\!(\theta_1\mu_1\!-\!\theta_2\mu_2)\frac{\partial\hat{E}}{\partial\xi}}{(\theta_1\mu_1\!+\!\theta_2\mu_2)\hat{E}}}\sim(1\!-\!\theta_j)[(2\!-\!\theta_j)\mathcal{M}-1]+ \frac{(1\!-\!\theta_j)(2\!-\!\theta_j)[1\!-\!(1\!-\!\theta_j+\theta_j^2)\mathcal{M}]\overline{M}|\xi|}{(1\!+\!\epsilon)\theta_j}
\nonumber\\
&&\quad-\frac{(1\!-\!\theta_j)^2\xi^2}{(1\!+\!\epsilon)\theta_j} \{1\!-\!\theta_j+(2\!-\!\theta_j)\mathcal{M}[4\!-5\theta_j\!+\!2\theta_j^2-(1\!+\! (1\!+\!\theta_j)^2)\mathcal{M}]\}, \quad\label{eqc16d}\\
&&\overline{\frac{M}{\theta_1\mu_1\!+\!\theta_2\mu_2}\!\int_0^\eta\!  \frac{\theta_1\mu_1\!-\!\theta_2\mu_2}{\theta_1\mu_1\!+\!\theta_2\mu_2}\frac{\partial\ln\hat{E}}{\partial\xi}}-\overline{\frac{M}{\theta_1\mu_1\!+\!\theta_2\mu_2}}\overline{\frac{M}{\theta_1\mu_1\!+\!\theta_2\mu_2}\!\int_0^\eta\!  \frac{\theta_1\mu_1\!-\!\theta_2\mu_2}{\theta_1\mu_1\!+\!\theta_2\mu_2}\frac{\partial\ln\hat{E}}{\partial\xi}}\nonumber\\
&&\sim\frac{(1\!-\!\theta_j)\overline{M}}{2}[3\!-\!2\theta_j-\!(2\!-\!\theta_j) \mathcal{M}] -\!\frac{(1\!-\!\theta_j)^2 (2\!-\!\theta_j)^2|\xi|}{\theta_j(1\!+\!\epsilon)}\!\!\left(\!\mathcal{M}\!-\!1\!+\!\frac{1\!-\!(1\!-\!\theta_j\!+\!\theta_j^2)\mathcal{M}}{2(1\!-\!\theta_j) (2\!-\!\theta_j)}\overline{M}^2\!\right)\!
\nonumber\\
&&+\frac{(1\!-\!\theta_j)^2\overline{M}\xi^2}{4\theta_j(1\!+\!\epsilon)}\left\{5\!-\!4\theta_j\!+\!2(2\!-\!\theta_j)\mathcal{M}[6-6\theta_j\!+\!2\theta_j^2-(1\!+\!(1\!-\!\theta_j)^2)\mathcal{M}]\right\} .
\label{eqc16e}
\end{eqnarray}
\end{subequations}

$\mathcal{B}$ involves $\Psi$, which also appears in Eq.~\eqref{eq15d}. We find
\begin{subequations}\label{eqc17}
\begin{eqnarray}
&&\!\Psi\sim\!\Theta(-\eta\!-\!|\xi|)\!\left[\mathcal{M}\frac{d\overline{M}}{d\xi}\frac{\mbox{erfc}\!\left(\!\frac{(1-\theta_j)|\xi|-\eta}{\sqrt{2(1\!+\!\epsilon)}}\!\right)\!}{\mbox{erfc}\!\left(\!\frac{(2-\theta_j)|\xi|}{\sqrt{2(1\!+\!\epsilon)}}\!\right)\!}-(1\!+\!\epsilon)\frac{\partial\hat{E}}{\partial\xi}\right]\!\!-\!2(1\!+\!\epsilon)\Theta(\eta\!+\!|\xi|)\!\left[ e^{-\frac{\eta+|\xi|}{w}} \frac{\partial\hat{E}}{\partial\eta}\mbox{sign}\xi\!\right. \nonumber\\
&&\left. \quad- \!\int_\eta^\infty\!\! e^{-\frac{\eta+|\xi|}{w}}\!\left(\frac{\partial^2\hat{E}}{\partial\xi\partial\eta}-\frac{\partial^2\hat{E}}{\partial\eta^2}\right)d\eta+\frac{1-\mathcal{M}}{2(1\!+\!\epsilon)} \frac{d\overline{M}}{d\xi}\frac{\mbox{erfc}\!\left(\!\frac{\eta-(1-\theta_j)|\xi|}{\sqrt{2(1\!+\!\epsilon)\theta_j}}\!\right)\!}{\mbox{erfc}\!\left(\!- \frac{(2-\theta_j)|\xi|}{\sqrt{2(1\!+\!\epsilon)\theta_j}}\!\right)\!} \right]\!,\label{eqc17a}
\end{eqnarray}
\begin{eqnarray}
&&\overline{\Psi\,\frac{\theta_1\mu_1\!-\!\theta_2\mu_2}{\theta_1\mu_1\!+\!\theta_2\mu_2}}\sim\frac{d\overline{M}}{d\xi}\left\{ \!\left[ \frac{(1\!-\!\theta_j)\mathcal{M} e^{-\frac{(2-\theta_j)^2\xi^2}{2(1+\epsilon)}}}{\mbox{erfc}\!\left(\!\frac{(2-\theta_j)|\xi|}{\sqrt{2(1\!+\!\epsilon)}}\!\right)\!}+\frac{(1\!-\!\mathcal{M})\sqrt{\theta_j}e^{-\frac{(2-\theta_j)^2\xi^2}{2(1+\epsilon)\theta_j}}}{\mbox{erfc}\!\left(\!-\frac{(2-\theta_j)|\xi|}{\sqrt{2(1\!+\!\epsilon)\theta_j}}\!\right)\!}\right]\mbox{sign}\xi \right.\nonumber\\
&&\quad\times\left.\sqrt{\frac{2(1\!+\!\epsilon)}{\pi}}-(2\!-\!\theta_j)(1\!-\!\theta_j\mathcal{M})\,\xi \right\}\! +(1\!-\!\theta_j)^2\!\left[\overline{M}+\frac{(2\!-\!\theta_j)^2 (\mathcal{M}\!-\!1)|\xi|}{\theta_j}\right]\!,\label{eqc17b}\\
&&\!\frac{\frac{\Psi}{(1+\epsilon)\hat{E}}}{\theta_1\mu_1\!+\!\theta_2\mu_2}\sim\frac{\Theta(-\eta\!-\!|\xi|)}{2}\!\left[\frac{\mathcal{M}}{1\!+\!\epsilon}\frac{d\overline{M}}{d\xi}\frac{\mbox{erfc}\!\left(\!\frac{(1-\theta_j)|\xi|-\eta}{\sqrt{2(1\!+\!\epsilon)}}\!\right)\!}{\mbox{erfc}\!\left(\!\frac{(2-\theta_j)|\xi|}{\sqrt{2(1\!+\!\epsilon)}}\!\right)\!\hat{E}\!}-\frac{\partial\ln\hat{E}}{\partial\xi}\right]\!\!+\frac{\Theta(\xi^2\!-\!\eta\!^2)}{\theta_j\hat{E}}\nonumber\\
&&\!\times\!\!\left[\!2\!\int_\eta^\infty\!\!e^{-\frac{\eta\!+\!|\xi|}{w}}\!\left(\frac{\partial^2\hat{E}}{\partial\xi\partial\eta}-\frac{\partial^2\hat{E}}{\partial\eta^2}\right)\! d\eta\!-\frac{1\!-\!\mathcal{M}}{1\!+\!\epsilon} \frac{d\overline{M}}{d\xi}\frac{\mbox{erfc}\!\left(\!\frac{\eta-(1-\theta_j)|\xi|}{\sqrt{2(1\!+\!\epsilon)\theta_j}}\!\right)\!}{\mbox{erfc}\!\left(\!- \frac{(2-\theta_j)|\xi|}{\sqrt{2(1\!+\!\epsilon)\theta_j}}\!\right)\!}\! - e^{-\frac{\eta+|\xi|}{w}} \frac{\partial\hat{E}}{\partial\eta}\mbox{sign}\xi\right]\! \nonumber\\
&&+\frac{\Theta(\eta\!-\!|\xi|)}{\theta_j\hat{E}}\!\!\left[2e^{\frac{\eta\!-\!2|\xi|}{w}}\!\int_\eta^\infty\!\!e^{-\frac{\eta}{w}}\!\left(\frac{\partial^2\hat{E}}{\partial\xi\partial\eta}-\frac{\partial^2\hat{E}}{\partial\eta^2}\right)\! d\eta\!- \frac{1\!-\!\mathcal{M}}{1\!+\!\epsilon} \frac{d\overline{M}}{d\xi}e^{\frac{\eta\!-\!|\xi|}{w}}\frac{\mbox{erfc}\!\left(\!\frac{\eta-(1-\theta_j)|\xi|}{\sqrt{2(1\!+\!\epsilon)\theta_j}}\!\right)\!}{\mbox{erfc}\!\left(\!- \frac{(2-\theta_j)|\xi|}{\sqrt{2(1\!+\!\epsilon)\theta_j}}\!\right)\!}\right.\nonumber\\
&&\left. - e^{-\frac{2|\xi|}{w}} \frac{\partial\hat{E}}{\partial\eta}\mbox{sign}\xi  \right]\!\!.\quad\quad\quad\label{eqc17c}\end{eqnarray}
As in the case of Eq.~\eqref{eqc12e}, only the first term of Eq.~\eqref{eqc17c} contributes to its integral on $(0,\eta)$: all other have orders $w$ and $e^{-2|\xi|/w}$. Then
\begin{eqnarray}
\!\int_0^\eta\!\!\frac{\frac{\Psi}{(1+\epsilon)\hat{E}}}{\theta_1\mu_1\!+\!\theta_2\mu_2}\!\sim\!\frac{\Theta(-\eta\!-\!|\xi|)}{2}\!\!\left[\frac{\frac{d\overline{M}}{d\xi}\,\frac{\mathcal{M}}{1\!+\!\epsilon}}{\mbox{erfc}\!\left(\!\frac{(2-\theta_j)|\xi|}{\sqrt{2(1\!+\!\epsilon)}}\!\right)\!}\!\int_{-|\xi|}^\eta\! \frac{\mbox{erfc}\!\left(\!\frac{(1-\theta_j)|\xi|-\eta}{\sqrt{2(1\!+\!\epsilon)}}\!\right)\! }{\hat{E}}d\eta\!-\!\!\int_{-|\xi|}^\eta\!\! \frac{\partial\ln\hat{E}}{\partial\xi}d\eta\right]\!\!.\quad
 \label{eqc17d}
\end{eqnarray}
\end{subequations}

Then the following terms appear in Eqs.~\eqref{eq15c} and \eqref{eq15d}:
\begin{subequations}\label{eqc18}
\begin{eqnarray}
&&\overline{\frac{M}{\theta_1\mu_1\!+\!\theta_2\mu_2}\!\int_0^\eta\!  \frac{\frac{\Psi}{(1\!+\!\epsilon)\hat{E}}\, d\eta}{\theta_1\mu_1\!+\!\theta_2\mu_2}}\!\sim\!\frac{\theta_1\!-\!\theta_2}{4}\overline{M}\!+\!\frac{2\!-\!\theta_j}{4}\!\left[\frac{(1\!-\!\theta_j)(2\!-\!\theta_j)(1\!+\!\mathcal{M})\xi}{\theta_j}+\frac{d\overline{M}}{d\xi}\mathcal{M}|\xi|\right]\!\nonumber\\
&&\quad-\sqrt{\frac{1\!+\!\epsilon}{2\pi}}\mathcal{M}\frac{d\overline{M}}{d\xi} \frac{e^{-\frac{(2-\theta_j)^2\xi^2}{2(1+\epsilon)}}}{2\,\mbox{erfc}\!\left(\frac{(2-\theta_j)|\xi|}{\sqrt{2(1+\epsilon)}}\right)\!}, \label{eqc18a}\\
&&\overline{\hat{E}\!\int_0^\eta\!  \frac{\frac{\Psi}{(1\!+\!\epsilon)\hat{E}}\, d\eta}{\theta_1\mu_1\!+\!\theta_2\mu_2}}\sim\!\mathcal{M}\frac{d\overline{M}}{d\xi} \frac{2g\!\left(\!\frac{(2-\theta_j)|\xi|}{\sqrt{2(1\!+\!\epsilon)}}\right)\!}{\mbox{erfc}\!\left(\!\frac{(2-\theta_j)|\xi|}{\sqrt{2(1\!+\!\epsilon)}}\right)\!}\!+\!\frac{\mathcal{M}}{4}(1\!-\!\theta_j)\mbox{sign}\xi\!-\!\frac{(1\!-\!\theta_j)(2\!-\!\theta_j)}{4\theta_j(1\!+\!\epsilon)}\nonumber\\
&&\,\quad\quad\quad\quad\quad\quad\quad \times[2(2\!-\!\theta_j)(\mathcal{M}-1)+\theta_j][(2\!-\!\theta_j)|\xi|\mathcal{M}-\overline{M}]\xi. \label{eqc18b}
\end{eqnarray}
\end{subequations}

Eqs.~\eqref{eq15c}, \eqref{eq15d}, \eqref{eqc16}, \eqref{eqc17} and \eqref{eqc18} yield
\begin{subequations}\label{eqc19}
\begin{eqnarray}
&&\!\mathcal{B}\sim(\theta_1\!-\!\theta_2)\overline{M}+\epsilon\frac{\theta_1\!-\!\theta_2}{1\!+\!\epsilon}\!\left\{\overline{M}\!\left[\frac{(\theta_1\!-\!\theta_2)(2\!-\!\mathcal{M})}{8}-\frac{(1\!-\!\theta_j)[3\!-\!2\theta_j-(2\!-\!\theta_j)\mathcal{M}]}{2}\!-\theta_1\theta_2\right] 
\right.\nonumber\\
&&+\sqrt{\frac{2(1\!+\!\epsilon)}{\pi}}\frac{d\overline{M}}{d\xi}\mbox{sign}\xi \!\left[ \frac{(1\!-\!\theta_j\!-\!\frac{1}{4}  \mbox{sign}\xi)\mathcal{M} e^{-\frac{(2-\theta_j)^2\xi^2}{2(1+\epsilon)}}}{\mbox{erfc}\!\left(\!\frac{(2-\theta_j)|\xi|}{\sqrt{2(1\!+\!\epsilon)}}\!\right)\!}+\frac{(1\!-\!\mathcal{M})\sqrt{\theta_j}e^{-\frac{(2-\theta_j)^2\xi^2}{2(1+\epsilon)\theta_j}}}{\mbox{erfc}\!\left(\!-\frac{(2-\theta_j)|\xi|}{\sqrt{2(1\!+\!\epsilon)\theta_j}}\!\right)\!}\right]\nonumber\\
&&-\overline{M}\frac{d\overline{M}}{d\xi}\mathcal{M} \frac{g\!\left(\frac{(2-\theta_j)|\xi|}{\sqrt{2(1+\epsilon)}}\right)\!}{\mbox{erfc}\!\left(\frac{(2-\theta_j)|\xi|}{\sqrt{2(1+\epsilon)}}\right)\!}+\frac{ (2\!-\!\theta_j)|\xi|}{\theta_j} \!\left[ \theta_j\mathcal{M}\frac{d\overline{M}}{d\xi}\!\left(\frac{1}{4}\!+\!\theta_j\right)\!+\! \frac{(1\!-\!\theta_j)\overline{M}^2}{2(1+\epsilon)} + (2\!-\!\theta_j)(1\!+\!\mathcal{M})\right.\nonumber\\
&&\left.\!\times\!\left(\frac{  (1\!-\!\theta_j) \mbox{sign}(\xi)}{4} \!-\!2\mathcal{M}\right)\!\!-2(1\!-\!\theta_j)^2(2\!-\!\theta_j)(1\!-\!\mathcal{M})\!-\!(1-\theta_j+\theta_j^2)\frac{\overline{M}^2}{2}\mathcal{M}\!\right]\nonumber\\
&&+\frac{(1\!-\!\theta_j)\overline{M}\xi^2}{4\theta_j(1\!+\!\epsilon)}\!\left[ [2(1\!-\!\theta_j)(1\!+\!(1\!-\!\theta_j)^2)+(2\!-\!\theta_j)^3\mbox{sign}\xi ]\mathcal{M}(\mathcal{M}\!-\!1) -(1\!-\!\theta_j)(5\!-\!4\theta_j)\right.\nonumber\\
&&\left.\left.+ \!\left(\frac{\theta_j\mbox{sign}\xi}{2}\!-\!2(1\!-\!\theta_j)(2\!-\!\theta_j)\!\right)\!(2\!-\!\theta_j)^2\mathcal{M}\right]\!\right\}\!.\quad\label{eqc19a}
\end{eqnarray}
As $|\xi|\gg 1$, Eq.~\eqref{eqc19a} becomes
\begin{eqnarray}
\mathcal{B}\sim (\theta_1-\theta_2)\sqrt{\frac{1\!+\!\epsilon}{2\pi\theta_j}} \, e^{-\frac{(2\!-\!\theta_j)^2\xi^2}{2(1\!+\!\epsilon)\theta_j}}\!\left[1-\frac{\epsilon(1\!-\!\theta_j)^2(5-4\theta_j)\xi^2}{4\theta_j(1+\epsilon)^2}\right]\!. \label{eqc19b}
\end{eqnarray}
\end{subequations}

\section{Stationary moments as $|\xi|\to\infty$} \label{ap:d}
In the limit as $|\xi|\to\infty$ the stationary probability density is given by Eqs.~\eqref{eq17} with the approximations \eqref{eq19} and \eqref{eq20a}. The latter yields the following average and variance
\begin{eqnarray*}
&&\langle\xi\rangle_s\sim\!\left(\sum_{j=1}^2\sqrt{\frac{7}{\theta_j}-5+\theta_j}\, \right)^{-1}\!\sqrt{\frac{2}{\pi\epsilon}}\sum_{j=1}^2\!(-1)^j\!\int_0^\infty\! \xi e^{-\frac{\theta_j\xi^2}{2\epsilon(7-5\theta_j +\theta_j^2)}}d\xi, \\
&&\langle\xi^2\rangle_s\sim\!\left(\sum_{j=1}^2\sqrt{\frac{7}{\theta_j}-5+\theta_j}\, \right)^{-1}\!\sqrt{\frac{2}{\pi\epsilon}}\sum_{j=1}^2\!\int_0^\infty\! \xi^2 e^{-\frac{\theta_j\xi^2}{2\epsilon(7-5\theta_j +\theta_j^2)}}d\xi, 
\end{eqnarray*}
respectively. By performing the integrals, we obtain Eqs.~\eqref{eq20c} and \eqref{eq20f}. The other moments involve integrals over $\eta$ and $\xi$. It is immediate to show
\begin{subequations}\label{eqd1}
\begin{eqnarray}
&&\langle\eta\rangle_s=\left\langle\eta-\frac{\theta_1-\theta_2}{2}\xi\right\rangle_s+\frac{\theta_1-\theta_2}{2}\langle\xi\rangle_s,\label{eqd1a}\\  
&&\langle\eta^2\rangle_s=\left\langle\eta\!\left(\eta-\frac{\theta_1-\theta_2}{2}\xi\right)\!\right\rangle_s+\frac{\theta_1-\theta_2}{2}\!\left\langle\xi\!\left(\eta-\frac{\theta_1-\theta_2}{2}\xi\right)\!\right\rangle_s\!+\!\left(\frac{\theta_1-\theta_2}{2}\right)^2\!\langle\xi^2\rangle_s.\label{eqd1b}
\end{eqnarray}
\end{subequations}

Eqs.~\eqref{eq19} produce
\begin{eqnarray}
\overline{\eta\!-\!\frac{\theta_1\!-\!\theta_2}{2}\xi}\sim\!\sqrt{\frac{2(1\!+\!\epsilon)}{\pi}}\!\sum_{j=1}^2\!\frac{\Theta_j}{\zeta_j}(1\!-\!\theta_j)e^{-\frac{(2\!-\!\theta_j)^2\xi^2}{2\theta_j(1+\epsilon)}}\!=\!\sqrt{\frac{1\!+\!\epsilon}{2\pi}}(\theta_1\!-\!\theta_2)\! \sum_{j=1}^2\!\frac{\Theta_j(-1)^je^{-\frac{(2\!-\!\theta_j)^2\xi^2}{2\theta_j(1+\epsilon)}}}{\zeta_j}\!.\quad \label{eqd2} 
\end{eqnarray}
Multiplying this expression by $\hat{F}(\xi)$ and integrating, we get
\begin{eqnarray}
\left\langle\eta-\frac{\theta_1-\theta_2}{2}\xi\right\rangle_s\sim \frac{\!\sqrt{\frac{1\!+\!\epsilon}{\epsilon}}(\theta_1\!-\!\theta_2)}{\pi\sum\sqrt{\frac{7}{\theta_j}-5+\theta_j}}\! \sum_{j=1}^2\!\int_0^\infty\!\frac{(-1)^j}{\zeta_j}e^{-\frac{\xi^2}{2\epsilon(\frac{7}{\theta_j}-5+\theta_j)}-\frac{(2-\theta_j)^2\xi^2}{2\theta_j(1+\epsilon)}}d\xi. \label{eqd3}
\end{eqnarray}
We now change variables in the integral to $x=\xi/\sqrt{2\epsilon(\frac{7}{\theta_j}-5+\theta_j)}$, thereby obtaining
\begin{eqnarray*}
\left\langle\eta\!-\!\frac{\theta_1-\theta_2}{2}\xi\right\rangle_s\!\sim\! \frac{\!\sqrt{2(1\!+\!\epsilon)}(\theta_1\!-\!\theta_2)}{\pi\sum\sqrt{\frac{7}{\theta_j}\!-\!5\!+\!\theta_j}}\! \sum_{j=1}^2\!(-1)^j\!\sqrt{\frac{7}{\theta_j}\!-\!5\!+\!\theta_j}\! \int_0^\infty\!\frac{e^{-x^2[1\!+\!\epsilon\frac{(2\!-\!\theta_j)^2(\frac{7}{\theta_j}\!-\!5\!+\!\theta_j)}{\theta_j(1+\epsilon)}]}dx}{\zeta_j\!\left(\sqrt{2\epsilon(\frac{7}{\theta_j}\!-\!5\!+\!\theta_j)}\,x\right)\!},\quad 
\end{eqnarray*}
from which
\begin{eqnarray}
\left\langle\eta\!-\!\frac{\theta_1-\theta_2}{2}\xi\right\rangle_s\!\sim\! \frac{\!\sqrt{1\!+\!\epsilon}(\theta_1\!-\!\theta_2)}{\sqrt{2\pi}\sum_{j=1}^2 \sqrt{\frac{7}{\theta_j}\!-\!5\!+\!\theta_j}}\! \sum_{j=1}^2\!\frac{(-1)^j}{1+\sqrt{\theta_j}}\sqrt{\frac{7}{\theta_j}\!-\!5\!+\!\theta_j}\,, \label{eqd4}
\end{eqnarray}
after substituting $\epsilon=0$ into the integrand. Together with Eq.~\eqref{eq20c}, Eqs.~\eqref{eqd1a} and \eqref{eqd3} produce Eq.~\eqref{eq20b}.

Similar algebra and use of Eq.~\eqref{eqd2} yield
\begin{eqnarray}
\left\langle\!\left(\eta\!-\!\frac{\theta_1-\theta_2}{2}\xi\right)\xi\right\rangle_s\!\sim\!\frac{\!\sqrt{\epsilon(1\!+\!\epsilon)}(\theta_1\!-\!\theta_2)}{\pi \sum_{j=1}^2 \sqrt{\frac{7}{\theta_j}\!-\!5\!+\!\theta_j}} \sum_{j=1}^2\frac{\frac{7}{\theta_j}\!-\!5\!+\!\theta_j}{1+\sqrt{\theta_j}}, \label{eqd5}
\end{eqnarray}
which produces Eq.~\eqref{eq20e}.

Lastly, we get form Eq.~\eqref{eq19},
\begin{eqnarray}
&&\overline{\eta\left(\eta\!-\!\frac{\theta_1\!-\!\theta_2}{2}\xi\right)}\sim\!\sum_{j=1}^2\!\frac{\Theta_j}{\zeta_j}\!\left\{\sqrt{\frac{2(1\!+\!\epsilon)}{\pi}}(1\!-\!\theta_j)|\xi|e^{-\frac{(2\!-\!\theta_j)^2\xi^2}{2\theta_j(1+\epsilon)}}\!+\!(1\!+\!\epsilon)\!\left[\!\theta_j^\frac{3}{2}\,\mbox{erfc}\!\left(\!\!-\frac{(2\!-\!\theta_j)|\xi|}{\sqrt{2(1\!+\!\epsilon)\theta_j}}\!\right)\right.\right.\nonumber\\
&&\quad\quad\quad\quad\quad\quad\quad\left.\left.+\, e^{\!-\frac{(1\!-\!\theta_j)(2\!-\!\theta_j)^2\!\xi^2}{2\theta_j(1\!+\!\epsilon)}}\mbox{erfc}\!\left(\!\frac{(2\!-\!\theta_j)|\xi|}{\sqrt{2(1\!+\!\epsilon)}}\!\right)\!\right]\!\right\}\!.\quad \label{eqd6} 
\end{eqnarray}
Since $\theta_j^\frac{3}{2}=\sqrt{\theta_j}+(\theta_j-1)\sqrt{\theta_j}$,
\begin{eqnarray}
\left\langle\!\eta\!\left(\eta\!-\!\frac{\theta_1\!-\!\theta_2}{2}\xi\right)\!\!\right\rangle_s\!\sim\! 1\!+\!\epsilon\!+\!\frac{(1\!+\!\epsilon)(\theta_1\!-\!\theta_2)}{\! \sum\!\sqrt{\frac{7}{\theta_j}\!-\!5\!+\!\theta_j}} \sum_{j=1}^2\!\frac{(-1)^j}{1\!+\!\!\sqrt{\theta_j}}\!\!\left[\frac{\sqrt{\epsilon}}{\!\sqrt{1\!+\!\epsilon}}\frac{\frac{7}{\theta_j}\!-\!5\!+\!\theta_j}{\pi}\!-\!\frac{\sqrt{7\!-\!5\theta_j\!+\!\theta_j^2}}{2}\right]\!\!.\quad\quad\label{eqd7}
\end{eqnarray}
Substitution of Eqs.~\eqref{eqd5} and \eqref{eqd7} into Eq.~\eqref{eqd1b} yields Eq.~\eqref{eq20d}.

\section{Limit as $|\xi|\to\infty$ for the evolution of the marginal density} \label{ap:e}
To describe the motion of wave fronts appearing in the marginal probability density we need to approximate certain integrals appearing in Section~\ref{sec:5}. For instance, as $|\Xi|\to\infty$, the integrand is dominated in Eq.~\eqref{eq23} by the exponential of $-\frac{\eta^2}{2(1+\epsilon)}-\frac{\eta+|\Xi|}{w}$. Splitting the integral into integrals over $-\infty<\eta<-|\Xi|$ and over $-|\Xi|<\eta<\infty$, the first of them dominates and a simple integration by parts \cite{bender} produces the approximation in Eq.~\eqref{eq23}. Consider the approximate solution of Eq.~\eqref{eq23} for positive large $\Xi$ and long times: 
\begin{eqnarray}
&&\frac{\tau}{\sqrt{\pi}}=\int^x e^{x^2}dx\sim\frac{e^{x^2}}{2x},\quad x=\frac{\Xi}{\sqrt{2(1+\epsilon)}}\Longrightarrow 
x^2=\frac{1}{2}\ln x^2+\ln(q\tau),\quad q=\frac{2}{\sqrt{\pi}}.\label{eqe1}
\end{eqnarray}
The method of dominant balance \cite{bender} for $\tau\gg 1$ produces $x^2=\ln(q\tau)+\delta$ with $\delta\ll\ln(q\tau)$ which, inserted into Eq.~\eqref{eqe1} gives
\begin{eqnarray*}
\delta=\frac{1}{2}\ln[\ln(q\tau)+\delta]=\frac{1}{2}\ln[\ln(q\tau)]+\frac{1}{2}\ln\!\left[1+\frac{\delta}{\ln(q\tau)}\right]\!\approx \frac{1}{2}\ln[\ln(q\tau)]+\frac{\delta}{2\ln(q\tau)}. 
\end{eqnarray*}
Then substituting $\delta\approx\frac{1}{2}\ln[\ln(q\tau)]$ into the relation $x^2=\ln(q\tau)+\delta$, we obtain
\begin{eqnarray}
x^2\approx  \ln(q\tau)+\frac{1}{2}\ln[\ln(q\tau)]= \ln(q\tau)\!\left(1+\frac{\ln[\ln(q\tau)]}{2\ln(q\tau)}\right)\!\Longrightarrow x\approx\sqrt{\ln(q\tau)}+\frac{1}{4}\frac{\ln[\ln(q\tau)]}{\sqrt{\ln(q\tau)}}, \quad\label{eqe2}
\end{eqnarray}
which is Eq.~\eqref{eq25}. In the case of different temperatures, a similar calculation with Eqs.~\eqref{eq36d} and \eqref{eq37} produce the relation $x^3 e^{-x^2}=1/(q\tau)$ with $q=(\theta_1-\theta_2)\theta_j/(2-\theta_j)/\sqrt{\pi}$. Then the method of dominant balance yields the approximation in Eq.~\eqref{eq38d}.
\end{widetext}

\section{Effect of different diode mobilities}\label{ap:f}
Let us consider a diode having a piecewise continuous current-voltage characteristics in nondimensional form:
\begin{subequations}\label{eqf1}
\begin{eqnarray}
I_D(u)=-\varepsilon\Theta(-\varepsilon-u)+ u\Theta(u+\varepsilon). \label{eqf1a}
\end{eqnarray}
Eq.~\eqref{eqf1a} is a piecewise linear approximation to the current of an ideal diode in series with a resistor. Here, $\varepsilon\ll 1$ is a small saturation current. The corresponding conductance is 
\begin{eqnarray}
\mu(u)=-\frac{\varepsilon}{u}\Theta(-\varepsilon-u)+ \Theta(u+\varepsilon). \label{eqf1b}
\end{eqnarray}
\end{subequations}
For $\theta_i=1$, $i=1,2$, (equal temperatures), we get
\begin{widetext}
\begin{eqnarray*}
&&\mathcal{A}=\frac{4\varepsilon}{\sqrt{2\pi(1+\epsilon)}}\left( \int_{-\infty}^{\varepsilon-\xi}\frac{e^{-\frac{\eta^2}{2(1+\epsilon)}}d\eta}{\xi-\eta+\varepsilon}+ \frac{\varepsilon}{2\xi}\int_{\varepsilon-\xi}^{\xi-\varepsilon} e^{-\frac{\eta^2}{2(1+\epsilon)}}d\eta+ \int_{\xi-\varepsilon}^\infty\frac{e^{-\frac{\eta^2}{2(1+\epsilon)}}d\eta}{\xi+\eta+\varepsilon} \right)\\
&&=\frac{4\varepsilon}{\sqrt{2\pi(1+\epsilon)}}\left( \int_{-\infty}^{\varepsilon-\xi}\frac{(\xi+\eta-\varepsilon)e^{-\frac{\eta^2}{2(1+\epsilon)}}d\eta}{2\xi(\xi-\eta+\varepsilon)}+ \frac{\varepsilon}{2\xi}\int_{-\infty}^\infty e^{-\frac{\eta^2}{2(1+\epsilon)}}d\eta+ \int_{\xi-\varepsilon}^\infty\frac{(\xi-\eta-\varepsilon)e^{-\frac{\eta^2}{2(1+\epsilon)}}d\eta}{2\xi(\xi+\eta+\varepsilon)} \right)\\
&&=\frac{2\varepsilon}{\xi}\left(1+\frac{2}{\sqrt{2\pi(1+\epsilon)}}\int_{\xi-\varepsilon}^\infty\frac{(\xi-\eta-\varepsilon)e^{-\frac{\eta^2}{2(1+\epsilon)}}d\eta}{\xi+\eta+\varepsilon} \right)\!,
\end{eqnarray*}
\end{widetext}
after a change of variable is carried out in the first integral. In the limit as $\xi\to\infty$, we obtain
\begin{subequations}\label{eqf2}
\begin{eqnarray}
\mathcal{A}\sim \frac{2\varepsilon}{\xi}\left(1-\frac{(1+\epsilon)^\frac{3}{2}}{\xi^2\sqrt{2\pi}} e^{-\frac{(\xi-\varepsilon)^2}{2(1+\epsilon)}}\right)\!.\label{eqf2a}
\end{eqnarray}
Thus, $\xi\mathcal{A}\sim 2\varepsilon$  in Eq.~\eqref{eq29} and the motion of the wave front occurs on a slow time scale $\varepsilon s$. Eq.~\eqref{eq32c} for the variance of the forefront becomes simply 
\begin{eqnarray}
\frac{d\sigma}{d\Xi}\sim  \frac{2}{\Xi}\Longrightarrow \sigma\sim 2\ln\Xi, \quad\mbox{as $\Xi\to\infty$.} \label{eqf2b}
\end{eqnarray}
\end{subequations}

\end{document}